\documentclass[aps,twocolumn,amsmath,amssymb,preprintnumbers]{revtex4}
\usepackage{amsmath} \usepackage{amsfonts} \usepackage{amssymb}
\usepackage{bbm}
\usepackage{epsfig}
\usepackage{graphics}
\usepackage{graphicx}
\newcommand{\be}{\begin{equation}}
\newcommand{\ee}{\end{equation}}
\newcommand{\ba}{\begin{eqnarray}}
\newcommand{\ea}{\end{eqnarray}}

\newcommand{\hmn}{_{\hat\mu\hat\nu}}

\newcommand{\iv}{^{(4)}}

\newcommand{\iD}{^{(D)}}
\newcommand{\hmnrs}{_{\hat\mu\hat\nu\hat\rho\hat\sigma}}
\newcommand{\nt}{^{(\textup{int})}}

\begin{document}

\title[ ]{Warping with dilatation symmetry and self-tuning of the cosmological constant}

\author{C. Wetterich}
\affiliation{Institut  f\"ur Theoretische Physik\\
Universit\"at Heidelberg\\
Philosophenweg 16, D-69120 Heidelberg}

\begin{abstract}
We investigate consequences of an ultraviolet fixed point in quantum gravity for the cosmological constant. For this purpose we perform dimensional reduction of a general dilatation symmetric effective action $\Gamma$  in dimension $d>4$ to an effective four-dimensional theory of gravity with a dilaton field. We find a stable flat phase in the space of extrema of $\Gamma$ which results in a vanishing four-dimensional cosmological constant $\Lambda$. In order to understand the self-tuning mechanism leading to $\Lambda = 0$ we discuss in detail the most general warped geometries with maximal four-dimensional symmetry and $SO(d-4)$ isometry of internal space. While the solutions of the $d$-dimensional field equations admit singular spaces with arbitrary $\Lambda$, the extremum condition for $\Gamma$ imposes additional restrictions which result in $\Lambda = 0$. In cosmology, the dilatation symmetric fixed point may only be reached for asymptotic time $t \to \infty$. At finite $t$ dilatation anomalies result in an effective potential and mass for the pseudo-dilaton or cosmon and in dark energy. 
\end{abstract}

\maketitle

\section{Introduction}
\label{intro}
The possible importance of a dilatation symmetric fixed point for the understanding of the cosmological constant problem has been stressed long ago \cite{CWQ},\cite{CWAA},\cite{CWA}. This fixed point may be associated with an ultraviolet fixed point of quantum gravity, as found \cite{26} in functional renormalization group studies of the asymptotic safety scenario \cite{25}. Indeed, a fixed point is typically visible not only in the running of couplings with varying momentum, but also for suitable varying background fields. The close correspondence between momentum flow and background field flow has been established first for four-dimensional scalar theories \cite{CoWei}, but holds in a much wider context. In the region of field space which is dominated by the fixed point the quantum effective action $\Gamma$ must be dilatation invariant, whereas close to the fixed point small "`dilatation anomalies"' govern the flow of couplings in the vicinity of the fixed point. We emphasize that these statements hold for the effective action $\Gamma$ which includes all effects of quantum fluctuations - the fixed point may typically even be generated by the quantum fluctuations.

The central idea for a solution of the cosmological constant problem \cite{CWQ} argues that the effective cosmological constant $\Lambda$ vanishes precisely at the fixed point. It is then sufficient that cosmological "`runaway solutions"' drive the fields into the region of the fixed point as time $t$ increases. For a cosmological runaway solution the values of fields are not static but continue to change for all times. Only for $t\to\infty$ the field equations derived from the fixed point effective action become accurate. Since such a fixed point is generally reached only asymptotically for $t\to\infty$, it is only in this limit that the field equations exhibit an exact dilatation symmetry. For finite $t$ the dilatation anomalies are still present, although their effects are small since they have to vanish for $t\to\infty$. Based on this scenario a dynamical form of homogeneous dark energy has been predicted long before observational discovery \cite{CWQ}. (See also the subsequent papers \cite{Q}.)

At the fixed point all memory of mass or length scales is lost and the quantum effective action becomes dilatation symmetric. In other words, the dilatation anomaly vanishes when it is evaluated for the field configurations corresponding to the fixed point \cite{CWCC}. In consequence, a scalar field becomes massless in the asymptotic limit, corresponding to the Goldstone boson of spontaneously broken dilatation symmetry. For the approach to the fixed point the dilatation anomaly is not yet zero, and correspondingly the scalar ``pseudo-Goldstone boson'' still has a small mass, that vanishes only asymptotically. These ideas are realized in practice in quintessence cosmologies, where the ``cosmon''-field plays the role of the pseudo-Goldstone boson of spontaneously broken anomalous dilatation symmetry \cite{CWQ}, \cite{CWCC}. The cosmon mass is varying with time and of the order of the Hubble parameter \cite{CWAA}.

The cosmological approach to a fixed point may also explain why the time variation of fundamental parameters as the fine structure constant or the electron to proton mass ratio is very small \cite{CWcoup}, \cite{CWA}. If such couplings run towards a nonzero fixed point in dependence of the time varying cosmon field moving to infinity for $t\to\infty$, the field-dependence and therefore the time dependence is expected to be small close to the fixed point. An interesting exception may be a relatively slow approach to a fixed point (or a weak instability) in the gauge singlet sector of physics beyond the standard model - this could result in time-varying neutrino masses and the scenario of "`growing neutrino quintessence"' \cite{CWN}.

It is a crucial ingredient for the validity of the fixed point scenario for the solution of the cosmological constant problem that the cosmological constant $\Lambda$ vanishes precisely at the fixed point. We have argued that a higher dimensional setting sheds new light on this issue \cite{CWCC}, \cite{CWNL}. In a series of two joint papers we therefore investigate higher dimensional models for which the quantum effective action $\Gamma$ exhibits an exact dilatation symmetry. Our central finding states that the cosmological constant problem is indeed solved if $\Gamma$ is dilatation symmetric. There are always extrema of $\Gamma$ for which $\Lambda$ vanishes. The remaining issue concerns the question if interesting particle physics obtains for those extrema.

The first paper \cite{AA} presents the general ideas and discusses the "`flat phase"' in the space of possible extrema of $\Gamma$ which results in a vanishing four-dimensional cosmological constant $\Lambda$. Simple  examples of higher dimensional geometries demonstrate that the flat phase can be realized for an almost arbitrary form of the dilatation symmetric effective action for a scalar-tensor-theory, provided that the fixed point effective action does not contain a non-polynomial potential for the scalar field. 

In the present paper we extend these findings by a discussion of the most general warped geometries with maximal four-dimensional symmetry and internal $SO(d-4)$ isometry. The solutions of the $d$-dimensional field equations admit arbitrary $\Lambda$ - the cosmological constant appears as an integration constant of the general solution. We show how the extremum conditions for $\Gamma$ restrict the allowed values of the integration constants and enforce a vanishing cosmological constant. We find that the existence of a ``flat phase'' of extrema of $\Gamma$ which lead to $\Lambda=0$ is a very general feature and discuss the associated ``self-tuning'' of the cosmological constant. The additional dimensions are crucial for this mechanism.

For the specific case of $SO(d-4)$ isometry we will find that the extremum conditions turn out to be so strong that all geometries with this symmetry lead to a diverging four-dimensional Planck mass. While this precludes the realization of $SO(d-4)$ symmetry together with four-dimensional Poincare-symmetry for a realistic compactification, the general conditions leading to $\Lambda = 0$ remain valid for much more general solutions with reduced symmetry, for which a finite Planck mass may be found.

The basic object of our investigation is the quantum effective action $\Gamma$ where all quantum fluctuations are already included. The field equations derived from $\Gamma$ are exact without any further quantum corrections. We do not postulate here that the effective action of a fundamental theory is dilatation symmetric - in general, it is not, due to dilatation anomalies. We only make the hypothesis that $\Gamma$ has a fixed point for certain asymptotic field values to be specified below. Only in this asymptotic limit the dilatation anomaly vanishes - the ``fixed point effective action'' is dilatation symmetric. We emphasize, nevertheless, that a dilatation symmetric effective action is also the starting point for approaches where dilatation symmetry is realized as an exact quantum symmetry \cite{SD}. (For early discussions of a dilatation symmetric standard model see \cite{Dil}, \cite{PSW}.)

For a pure gravity theory in $d$ dimensions the most general dilatation symmetric effective action takes the form
\be\label{A1}
\Gamma = \int_{\hat x} \hat g^{1/2}F(\hat R\hmnrs),
\ee
with $ \int_{\hat x}= \int d^d x$ and $\hat g = - \det(\hat g_{\hat \mu \hat \nu})$. Here $F$ transforms as a scalar under general coordinate transformations and has therefore to be constructed from the curvature tensor $\hat R\hmnrs$ and its covariant derivatives, contracted with appropriate combinations of the inverse metric $\hat g^{\hat \mu \hat \nu}$. (We use hats for $d$- dimensional objects and indices.) Dilatation symmetry forbids the use of parameters with dimension of mass or length in the construction of $F$. More generally, the effective action may also depend on fields other than gravity. In this paper we first concentrate on gravity coupled to a higher dimensional scalar dilaton field and turn back to the action \eqref{A1} only in sect. \ref{gravitywithout}.

We can write the most general form of a dilatation symmetric quantum effective action as
\begin{equation}\label{2A}
\Gamma=\int_{\hat x}\hat g^{1/2}{\cal L}.
\end{equation}
where ${\cal L}=F$ in the case of pure gravity. Dilatation transformations correspond to a rescaling of the metric by a constant factor $\alpha^2$,
\begin{eqnarray}\label{2B}
\hat g_{\hat\mu\hat\nu} \to \alpha^2\hat g_{\hat\mu\hat\nu}~,~\hat g^{1/2}\to\alpha^d\hat g^{1/2},
{\cal L}\to\alpha^{-d}{\cal L}.
\end{eqnarray}
A dilatation symmetric effective action remains invariant under these rescalings. This requirement also fixes the scaling of an additional $d$-dimensional dilaton field. 

The special role of dilatation symmetry for the existence of a flat phase of solutions with a vanishing cosmological constant is visible already for the most general form of a dilatation symmetric effective action. We are interested in solutions with a block diagonal metric
\begin{equation}\label{2c}
\hat g_{\hat\mu\hat\nu} (x,y)=
\left(\begin{array}{ccc}
\sigma(y)g^{(4)}_{\mu\nu} (x)
&,&0\\
0&,&g^{(D)}_{\alpha\beta}(y)
\end{array}\right)
\end{equation}
with $x^\mu$ the four-dimensional coordinates and $y^\alpha$ coordinates of internal space, with corresponding metrics $g^{(4)}_{\mu\nu}$ and $g^{(D)}_{\alpha\beta},d=D+4$. The function $\sigma(y)$ accounts for a possible warping \cite{RSW,7A,RDW,RS}. With 
$\hat g^{1/2}=(g^{(4)})^{1/2}\sigma^2(g^{(D)})^{1/2}$ we define
\begin{equation}\label{2D}
W(x)=\int_y(g^{(D)}{(y)})^{1/2}\sigma^2(y){\cal L}(x,y),
\end{equation}
such that
\begin{equation}\label{2E}
\Gamma=\int_x(g^{(4)})^{1/2} W,
\end{equation}
The scaling under dilatations $(g^{(4)}_{\mu\nu}\to\alpha^2 g^{(4)}_{\mu\nu}~,~g^{(D)}_{\alpha\beta}\to\alpha^2 g^{(D)}_{\alpha\beta})$
implies
\begin{equation}\label{2F}
W\to\alpha^{-4}W.
\end{equation}

Every extremum of $W$ realizes an extremum of $\Gamma$ in the flat phase \cite{AA} and implies $\Lambda=0$. Indeed, an extremum of $W$ can only occur for field configurations for which $W$ vanishes, $W_0=0$. This follows from the use of a neighboring rescaled field configuration \eqref{2B} with $\alpha = 1 + \epsilon.$ For an extremum of $W$ the variation with $\epsilon$ has to vanish, $\partial_\epsilon (1+\epsilon)^{-4} W_0=0,\text{thus}~ W_0 = 0$. In consequence the variation of $\Gamma$ also vanishes
\be\label{8A}
\delta \Gamma = \int_{\hat x} (\hat g_0^{1/2} \delta W + \delta \hat g^{1/2} W_0) = 0, 
\ee
realizing an extremum of $\Gamma$. This extremum occurs for $\Gamma_0 = 0$, such that for any effective four-dimensional theory this results in $\Lambda =0$. This follows for solutions with maximal four-dimensional symmetry where 
\be\label{8B}
\Gamma_0 = - \int_{x} (g^{(4)})^{1/2} \chi^2 \Lambda,
\ee
with $\chi$ the effective four-dimensional reduced Planck mass $(\chi ^2\, \widehat {=}\,  M_p^2 / 8 \pi)$. This property of the flat phase is very general and does not change if we add additional fields to the dilatation invariant effective action as, for example, a scalar $\xi$ in a scalar-tensor theory.

The paper \cite{AA} has focused on two issues:
\begin{itemize}
\item [(i)] The existence of extrema of $W$ for a large class of dilatation symmetric effective actions.
\item [(ii)] The dimensional reduction to effective four-dimensional gravity  and the establishment that $\Lambda=0$ in the flat phase.
\end{itemize}
Furthermore, a demonstration that an effective action which admits a flat phase does generically not allow other extrema with arbitrarily small $|\Lambda|$ is given in ref. \cite{CWNL}. In particular, there are no continuous families of extrema where $\Lambda$ appears as a continuous parameter. This issue is important for warped geometries with singularities where the existence of families of solutions of the higher dimensional field equations with continuous $\Lambda$ is known \cite{RSW}, \cite{7A}, \cite{RDW}. We will see that the extremum conditions for $\Gamma$ go beyond the higher dimensional field equations \cite{CWCON}. They precisely select the solutions with $\Lambda=0$ out of the continuous family of solutions.

The investigations of this paper are mainly based on the simplest possible dilatation symmetric effective action for a scalar-tensor theory, but the results are much more general as we argue in sects. \ref{generaldilatation}, \ref{gravitywithout}.
In the course of our discussion we will explicitly address the issues of quantum corrections, ``tuning of the cosmological constant to zero'', and ``naturalness'' of the solutions with $\Lambda=0$. Many general aspects are already discussed in \cite{CWCC}, \cite{AA} and not repeated here, such that we concentrate more on specific solutions. We find rather satisfactory answers to the naturalness problem. Asymptotic dilatation symmetry in higher dimensional theories may indeed provide the key for a solution of the cosmological constant problem. 

Our paper is organized as follows: In sect. \ref{dilatationsymmetric} we start with a simple dilatation symmetric effective action for a scalar-tensor theory and perform a convenient Weyl scaling. Sect. \ref{warped} discusses the most general form of the quasistatic solutions with $SO(D)$ isometry (as well as particular solutions with reduced isometry). This class of solutions shows the cosmological constant $\Lambda$ as a continuous free integration constant, together with other free integration constants. The general solution has up to two singularities. The properties of the metric close to the singularities are investigated in sect. \ref{structure}. The presence of the scalar field allows for a richer spectrum of possibilities than for pure gravity \cite{RSW,RDW,CWCON}. Sect. \ref{warpedsolutions} establishes criteria for obtaining an effective four-dimensional gravity, including cases where the volume of the $D$-dimensional subspace is infinite \cite{CWCRIT}, \cite{RS}. They are applied to select solutions which are acceptable from this point of view. In sect. \ref{geometry} we classify the singular solutions into ``zerowarp solutions'', first discussed in \cite{RSW} and ``warped branes'', first investigated in \cite{7A}. A discussion of the global geometrical properties in sect. \ref{global} closes the investigation of the most general solutions with $SO(D)$ isometry.

The second part of the paper deals with dimensional reduction to effective four-dimensional gravity. In sect. \ref{dimensional} we perform the reduction for the warped geometries with $SO(D)$ symmetry. The effective four-dimensional action depends on the integration constants which appear in the most general solution of the $d$-dimensional field equations. We show in sect. \ref{potentialfor} that the requirement that acceptable solutions should also obey the four-dimensional field equations severely restricts the allowed integration constants. Indeed, the four-dimensional field equations are more restrictive than the $d$-dimensional field equations since they reflect the extremum condition for arbitrary field variations which are local in four-dimensional space, but not necessarily in $d$-dimensional space. (The $d$-dimensional field equations correspond to variations which are local in $d$-dimensional space.) In particular, we find that the additional constraints leave $\Lambda=0$ as the only possibility.

We extend our arguments to a much more general form of the dilatation symmetric effective action in sect. \ref{generaldilatation}. We present the general argument that stable compactifications with finite four-dimensional gravitational constant and $\xi\neq 0$ must have $\Lambda=0$. For these extrema of $\Gamma$ both the dilaton and the radion have a vanishing mass. The radion corresponds to a rescaling of the characteristic length $l$ of internal space while keeping $\xi$ fixed. One may ask if the small mass of the radion away from the fixed point could turn it into a dark matter candidate.

In sect. \ref{gravitywithout} we discuss pure higher-dimensional gravity theories in the light of these findings. We show that a flat phase with $\Lambda = 0$ still exists in the absence of the $d$-dimensional dilaton field $\xi$. The role of the four-dimensional dilaton is now played by $l$ and coincides with the radion. We briefly address the possibility of ''geometrons'' - four-dimensional scalar fields with a flat potential in the limit of a dilatation symmetric $\Gamma$. They correspond to deformations of internal geometry within the space of extrema of $\Gamma$ which keep the characteristic length scale $l$ for internal geometry fixed. For the cosmology of the present epoch, which has not yet reached the fixed point, the geometrons acquire a small mass and one may again speculate about possible candidates for dark matter. 

Our conclusions in sect. \ref{conclusions} discuss the self-tuning of the cosmological constant to zero in the higher dimensional context. Two appendices investigate the special case of warped branes for five-dimensional dilatation symmetry and explore the possible class of $D$-dimensional hyperbolic Einstein spaces with finite volume and isometry $SO(D_1+1)~\times~SO(D-D_1)$.

\section{Dilatation symmetric gravity in higher dimensions}
\label{dilatationsymmetric}

Our starting point is the $d$-dimensional dilatation-symmetric action
\begin{equation}\label{1}
\Gamma=\int\hat g^{1/2}\left\{-\frac12\xi^2\hat R+\frac\zeta 2\partial^{\hat\mu}\xi\partial_{\hat\mu}\xi\right\}.
\end{equation}
This is the simplest form of the dilatation symmetric effective action in the presence of a scalar field $\xi$. Dilatations involve a rescaling of $\xi$,

\be\label{A12a}
\xi \to \alpha^{-\frac{d-2}{2}} \xi,
\ee
in addition to the rescaling of the metric \eqref{2B}. One may speculate that it corresponds to an exact ``ultraviolet'' fixed point in arbitrary $d$ in the limit $\xi\to\infty$. The existence of such a fixed point could be related to the formulation of a consistent quantum  gravity. Corrections to eq. \eqref{1} would then correspond to deviations from the fixed point. Besides diffeomorphism symmetry and dilatation symmetry the action \eqref{1} exhibits the extended scaling of the field equations discussed in \cite{AA}. We observe that for fixed momenta we obtain a free theory in the limit $\xi\to\infty$ if we expand around flat space. This adds to the plausibility for the existence of such a fixed point. 

\medskip\noindent
{\bf 1. Field equations}

The field equations obtain by variation of the effective action \eqref{1} with respect to the dilaton $\xi$ and the metric $\hat g_{\hat\mu\hat\nu}$, 
\begin{equation}\label{2}
\zeta\hat D^2\xi+\hat R\xi=0,
\end{equation}
and
\begin{eqnarray}\label{2a}
&&\xi^2\left(\hat R_{\hat\mu\hat\nu}-\frac12\hat R\hat g_{\hat\mu\hat\nu}\right)\\
&&=T^{(\xi)}_{\hat\mu\hat\nu}
=\zeta\partial_{\hat \mu}\xi\partial_{\hat\nu}\xi-\frac\zeta 2\partial^{\hat\rho}\xi\partial_{\hat\rho}\xi~\hat g_{\hat\mu\hat\nu}
+D_{\hat\nu}D_{\hat \mu}\xi^2-\hat D^2\xi^2\hat g_{\hat\mu\hat\nu}.\nonumber\label{3}
\end{eqnarray}
Here $D_{\hat \mu}$ is the $d$-dimensional covariant derivative, $\hat D^2=D^{\hat\mu}D_{\hat \mu}$ and $\hat g=-\det(\hat g_{\hat\mu\hat\nu})$. We are interested in cosmological solutions where the ``ordinary'' four space-time dimensions (with coordinates $x^\mu,\mu=0,\dots,3)$ and internal space (with coordinates $y^\alpha,\alpha=1,\dots,D)$ play a different role. (Higher dimensional indices $\hat\mu$ run from zero to $d-1,d=D+4.)$ Contracting eq. \eqref{2a} yields
\begin{equation}\label{3A}
\xi^2\hat R=\zeta\partial^{\hat\mu}\xi\partial_{\hat\mu}\xi+
\frac{2(d-1)}{d-2}
\hat D^2\xi^2
\end{equation}
and therefore
\begin{equation}\label{3B}
\Gamma=-\frac{d-1}{d-2}\int \partial_{\hat\rho}\{\hat g^{1/2}\partial^{\hat\rho}\xi^2\}.
\end{equation}

For all solutions with a vanishing boundary term the action vanishes. According to our discussion in the introduction this implies a vanishing four-dimensional cosmological constant for any extremum of $\Gamma$ which admits a dimensionally reduced effective four-dimensional theory. A simple solution of the field eqs. \eqref{2}, \eqref{2a} is
\begin{equation}\label{70AA}
\hat R_{\hat\mu\hat\nu}=0~,~\xi=\xi_0.
\end{equation}

In particular, a geometry ${\cal M}^4\times T^D$, with ${\cal M}^4$ flat four-dimensional Minkowski space and $T^D$ a $D$-dimensional torus with finite volume $\Omega_D$, solves the field equations and corresponds to an extremum of $\Gamma$. Dimensional reduction leads to consistent four-dimensional gravity with a nonzero effective Planck mass and vanishing effective cosmological constant $\Lambda=0$. This demonstrates that solutions in the flat phase always exist for the effective action \eqref{1}.

\medskip\noindent
{\bf 2. Weyl scaling}

For a systematic discussion of solutions of the field equations it is convenient to perform a coordinate change in field space by a Weyl scaling
\begin{equation}\label{4}
\hat g_{\hat\mu\hat\nu} =w^2\tilde g_{\hat\mu\hat\nu}.
\end{equation}
This results in
\begin{equation}\label{5}
\hat R=w^{-2}(\tilde R-f_d\partial^{\hat\mu}\ln w~\partial_{\hat \mu} \ln w-g_d
\hat D^2\ln w),
\end{equation}
and $f_d,g_d$ constants depending on $d$. Insertion into the action yields
\begin{eqnarray}\label{6}
\Gamma&=&\frac12\int\tilde g^{1/2}w^{d-2}\xi^2\\
&&\{-\tilde R+f_d\partial^{\hat \mu}\ln w~\partial_{\hat \mu}\ln w
+g_d\hat D^2\ln w\nonumber\\
&&+\zeta\partial^{\hat\mu}\ln\xi~\partial_{\hat \mu}\ln\xi\}.\nonumber
\end{eqnarray}
The indices are now raised and lowered with the metric $\tilde g_{\hat\mu\hat\nu}$, and this metric is also used for the connection in the covariant derivatives. Also $\tilde R$ is the curvature scalar computed from $\tilde g_{\hat\mu\hat\nu}$. Choosing
\begin{equation}\label{7}
w=M_d\xi^{-\frac{2}{d-2}}
\end{equation}
and omitting the total derivative $\sim g_d$
one obtains
\begin{equation}\label{8}
\Gamma=\frac{M^{d-2}_d}{2}
\int\tilde g^{1/2}\left\{-
\tilde R+\left(\zeta+\frac{4f_d}{(d-2)^2}\right)\partial^{\hat\mu}\ln\xi~\partial_{\hat \mu} \ln\xi\right\}.
\end{equation}
We can associate $M_d$ with the $d$-dimensional Planck mass. Using a rescaled scalar field
\begin{equation}\label{9a}
\delta=\left(\zeta+\frac{4f_d}{(d-2)^2}\right)^{1/2}\ln(\xi/M_d^{(d-2)/2})
\end{equation}
we arrive at
\begin{equation}\label{10a}
\Gamma=\frac{M^{d-2}_d}{2}\int\tilde g^{1/2}
\{-\tilde R+\partial^{\hat\mu}\delta~\partial_{\hat\mu}\delta\}.
\end{equation}
This action contains a higher dimensional Einstein term and the kinetic term for a free scalar field. A multiplicative scaling of $\xi$ corresponds to a constant shift in $\delta$. Thus $\delta$ is the Goldstone boson corresponding to spontaneously broken dilatation symmetry in a nonlinear field basis.

The field equations derived from the action \eqref{10a},
\begin{eqnarray}\label{11aneu}
\hat D^2\delta&=&0,\\
\tilde R_{\hat\mu\hat\nu}-\frac12\tilde R\tilde g_{\hat\mu\hat\nu}&=&\partial_{\hat \mu}\delta~\partial_{\hat\nu}\delta\label{12}
-\frac12\partial^{\hat\rho}\delta~\partial_{\hat\rho}\delta~\tilde g_{\hat\mu\hat\nu},
\end{eqnarray}
are exactly equivalent to the field equations \eqref{2}, \eqref{3}. We have only made a change of variables \eqref{4}, \eqref{7}, \eqref{9a}. They are, however, much easier to solve. In particular, the constants $\zeta$ and $M_d$ do not appear anymore - they influence only the translation to $\hat g_{\hat\mu\hat\nu}$ and $\xi$. Contracting the second equation yields
\begin{equation}\label{13}
\tilde R=\partial^{\hat\mu}\delta~\partial_{\hat\mu}\delta=D_{\hat\mu}(\delta\partial^{\hat\mu}\delta)
=\frac12\hat D^2\delta^2,
\end{equation}
where the last two identities use eq. \eqref{11aneu}. For all solutions of the field equations the action \eqref{10a} vanishes. 

An obvious solution of the field equations is 
\begin{equation}\label{13A}
\delta=const,\quad \tilde R_{\hat\mu\hat\nu}=0,
\end{equation}
corresponding to $\xi=const,~\hat R_{\hat\mu\hat\nu}=0$. The corresponding geometry could be $d$-dimensional Minkowski space, or a direct product of four-dimensional Minkowski space and a Ricci-flat $D$-dimensional internal space. If internal space has finite volume $\Omega_D$ the effective four-dimensional action is well defined. It has indeed a vanishing cosmological constant and a four-dimensional reduced Planck mass $M$ given by $M^2=M^{d-2}_d\Omega_D$. For a Ricci-flat space with non-abelian symmetry the four-dimensional theory will exhibit a non-abelian gauge symmetry. For an arbitrary isometry of internal space the four-dimensional gauge coupling is finite and nonzero. We want to learn more about the general solution and consider next geometries with warping and a possible non-uniform $\xi$ or $\delta$. 

\section{Warped solutions}
\label{warped}

In this and the next sections we discuss possible solutions for the ansatz
\begin{eqnarray}\label{14}
&&\tilde g_{\hat\mu\hat\nu}(x,y)=\left(\begin{array}{crcrl}
\sigma(z)\tilde g^{(4)}_{\mu\nu}(x)&,&0&,&\quad 0\\
0&,&L^2&,&\quad 0\\
0&,&0&,&L^2\rho(z)\bar g_{\bar\alpha\bar\beta}(\bar y)
\end{array}
\right),\nonumber\\
&&\delta(x,y)=\delta(z).
\end{eqnarray}
Here $D-1$ coordinates $\bar y^{\bar\alpha}$ form a homogeneous space,
\be\label{28A}
\bar R_{\bar\alpha\bar\beta}=Cg_{\bar\alpha\bar\beta}.
\ee

The internal coordinates $\bar y^{\bar\alpha}$ and $z$ are dimensionless and $L$ denotes the characteristic linear size of internal space. The four-dimensional metric $\tilde g^{(4)}_{\mu\nu}$ is used to define the four-dimensional Ricci-tensor $\tilde R^{(4)}_{\mu\nu}$ and we assume
\begin{equation}\label{15}
\tilde R^{(4)}_{\mu\nu}=\epsilon\tilde L^{-2}\tilde g^{(4)}_{\mu\nu}~,~\epsilon =\pm 1.
\end{equation}
The limit $\tilde L\to\infty$ corresponds to flat four-dimensional space with a vanishing cosmological constant, while finite $\tilde L$ with $\epsilon=1$ or $\epsilon=-1$ can describe four-dimensional de Sitter or anti-de Sitter space, with positive or negative cosmological constant, respectively. 

\medskip\noindent
{\bf 1. Warped field equations}

If we choose for $\bar g_{\bar\alpha\bar\beta}(\bar y)$ the metric of a $D-1$-dimensional hypersphere, our ansatz exhibits an $SO(D)$-isometry. Furthermore, we may assume that $\tilde g^{(4)}_{\mu\nu}(x)$ describes a four-dimensional space with maximal symmetry. In this case our ansatz describes the most general metric which is consistent with these symmetries. It is therefore well suited for a systematic study of all possible solutions that share $SO(D)$ and maximal four-dimensional symmetry. Still, it contains enough freedom to account for warping and a non-constant $\delta(z)$. For given constants $L,\tilde L,\epsilon$ we have to solve three differential equations for the three dimensionless functions $\sigma(z),\rho(z),~\delta(z)$, which depend on the dimensionless coordinate $z$. The function $\sigma(z)$ is the warp factor.

Let us start with the scalar field equation
\begin{eqnarray}\label{16}
\hat D^2\delta&=&\tilde g^{-1/2}\partial_{\hat \mu}(\tilde g^{1/2}\partial^{\hat \mu}\delta)\nonumber\\
&=&\sigma^{-2}\rho^{-\frac{D-1}{2}}\partial_z(\sigma^2\rho^{\frac{D-1}{2}}\partial_z\delta)L^{-2}=0.
\end{eqnarray}
Denoting $z$-derivatives by primes this reads
\begin{equation}\label{17}
\delta''+2\frac{\sigma'}{\sigma}\delta'+\frac{D-1}{2}\frac{\rho'}{\rho}\delta'=0.
\end{equation}
The general solution is
\begin{equation}\label{18}
\delta'=E\sigma^{-2}\rho^{-\frac{D-1}{2}},
\end{equation}
with $E$ a free integration constant.

For the gravitational field equations one has \cite{CWCC},\cite{CWCON}
\begin{eqnarray}\label{19}
&&2\tilde R_{\mu\nu}-\tilde R\tilde g_{\mu\nu}=L^{-2}\left\{-2\epsilon\left(\frac{L}{\tilde L}\right)^2\sigma^{-1}-
(D-1)C\rho^{-1}\right.\nonumber\\
&&\qquad\quad+3\frac{\sigma''}{\sigma}+\frac32(D-1)\frac{\rho'}{\rho}\frac{\sigma'}{\sigma}+(D-1)
\frac{\rho''}{\rho}+\\
&&\qquad\quad\left.\frac14(D-1)(D-4)
\left(\frac{\rho'}{\rho}\right)^2\right\}\tilde g_{\mu\nu}
=-L^{-2}\delta'^2\tilde g_{\mu\nu},\nonumber\\
&&2\tilde R_{\bar\alpha\bar\beta}-\tilde R\tilde g_{\bar\alpha\bar\beta}=L^{-2}
\left\{-4\epsilon\left(\frac{L}{\tilde L}\right)^2\sigma^{-1}-(D-3)C\rho^{-1}
\right.\nonumber\\\label{20}
&&\qquad+4\frac{\sigma''}{\sigma}+\left(\frac{\sigma'}{\sigma}\right)^2+2(D-2)\frac{\rho'}{\rho}
\frac{\sigma'}{\sigma}+(D-2)\frac{\rho''}{\rho}\\
&&\qquad\left.+\frac14(D-2)(D-5)\left(\frac{\rho'}{\rho}\right)^2\right\}\tilde g_{\bar\alpha\bar\beta}
=-L^{-2}\delta'^2\tilde g_{\bar\alpha\bar\beta},\nonumber\\
&&2\tilde R_{zz}-\tilde R\tilde g_{zz}=L^{-2}\left\{-4\epsilon\left(\frac{L}{\tilde L}\right)^2\sigma^{-1}
-(D-1)C\rho^{-1}\right.\nonumber\\\label{21}
&&\qquad\quad+3\left(\frac{\sigma'}{\sigma}\right)^2+2(D-1)\frac{\rho'}{\rho}\frac{\sigma'}{\sigma}\\
&&\qquad\quad\left.+\frac14(D-1)(D-2)\left(\frac{\rho'}{\rho}\right)^2\right\}
\tilde g_{zz}=L^{-2}\delta'^2\tilde g_{zz}.\nonumber
\end{eqnarray}
Only three of the four equations \eqref{17}, \eqref{19}-\eqref{21} are independent. Subtracting eq. \eqref{19} from eq. \eqref{20} one finds
\begin{eqnarray}\label{22}
\delta'^2&=&-C\rho^{-1}+\left(\frac{\sigma'}{\sigma}\right)^2-2\frac{\sigma''}{\sigma}\nonumber\\
&&+\frac{\rho'}{\rho}\frac{\sigma'}{\sigma}+\frac12(D-2)\left(\frac{\rho'}{\rho}\right)^2-
\frac12(D-2)\frac{\rho''}{\rho},
\end{eqnarray}
and the difference between eqs. \eqref{21} and \eqref{19} yields
\begin{eqnarray}\label{23}
\delta'^2&=&-\epsilon\left(\frac{L}{\tilde L}\right)^2\sigma^{-1}
+\frac32\left(\frac{\sigma'}{\sigma}\right)^2-\frac32\frac{\sigma''}{\sigma}+\frac14(D-1)
\frac{\rho'}{\rho}\frac{\sigma'}{\sigma}\nonumber\\
&&+\frac14(D-1)\left(\frac{\rho'}{\rho}\right)^2-\frac12(D-1)\frac{\rho''}{\rho}.
\end{eqnarray}
On the other hand, the derivative of eq. \eqref{21} implies
\begin{eqnarray}\label{24}
\delta''\delta'&=&2\epsilon\left(\frac{L}{\tilde L}\right)^2\frac{\sigma'}{\sigma^2}
+\frac12(D-1)C\frac{\rho'}{\rho^2}\nonumber\\
&&+3\left(\frac{\sigma'}{\sigma}\right)^2
\left(\frac{\sigma''}{\sigma'}-\frac{\sigma'}{\sigma}\right)\nonumber\\
&&+(D-1)\frac{\rho'}{\rho}\frac{\sigma'}{\sigma}
\left(\frac{\rho''}{\rho'}-\frac{\rho'}{\rho}+\frac{\sigma''}{\sigma'}-\frac{\sigma'}{\sigma}\right)\nonumber\\
&&+\frac14(D-1)(D-2)\left(\frac{\rho'}{\rho}\right)^2
\left(\frac{\rho''}{\rho'}-\frac{\rho'}{\rho}\right)\nonumber\\
&=&-2\frac{\sigma'}{\sigma}\delta'^2-\frac{D-1}{2}\frac{\rho'}{\rho}\delta'^2,
\end{eqnarray}
where the second identity uses eq. \eqref{17}. Multiplying eq. \eqref{23} with $-2\sigma'/\sigma$ and eq. \eqref{22} with $-(D-1)\rho'/(2\rho)$, and taking the sum yields eq. \eqref{24}. For a solution it is therefore sufficient to obey two suitable linear combinations of eqs. \eqref{19}-\eqref{21} and the scalar field equation \eqref{17}. 

Once $\delta'$ is expressed in terms if $\sigma$ and $\rho$ by eq. \eqref{18}, we therefore end with two second order differential equations for $\sigma$ and $\rho$. The local solution around some point $z_0$ has four integration constants, $\sigma(z_0),\sigma'(z_0),\rho(z_0),\rho'(z_0)$. They are not independent, however, since eq. \eqref{21} yields a relation between the four integration constants which depends on $\epsilon(L/\tilde L)^2$. In other words, we may take
\begin{equation}\label{25}
\Lambda=\epsilon\left(\frac{L}{\tilde L}\right)^2
\end{equation}
as one of the integration constants. For fixed $\Lambda$, eq. \eqref{21} constitutes a relation between $\sigma(z_0),\sigma'(z_0),\rho(z_0)$ and $\rho'(z_0)$, and only three further integration constants are left. For fixed $\Lambda$ it is also obvious that the solutions of the system of equations \eqref{19}-\eqref{21} are independent of $L$. 

\medskip\noindent
{\bf 2. Reduction of field equations}

One can express the derivative $\rho'/\rho$ in terms of $\sigma$. For this purpose we use eq. \eqref{21} and the combination of eqs. \eqref{19}, \eqref{20} that does not contain a term $\sim\rho''$, i.e.
\begin{eqnarray}\label{26}
\delta'^2&=&2D\Lambda\sigma^{-1}-(D-1)C\rho^{-1}-(D+2)
\frac{\sigma''}{\sigma}\nonumber\\
&&-(D-1)\left(\frac{\sigma'}{\sigma}\right)^2-\frac12(D-1)(D-2)\frac{\rho'}{\rho}\frac{\sigma'}{\sigma}\nonumber\\
&&+\frac14(D-1)(D-2)
\left(\frac{\rho'}{\rho}\right)^2.
\end{eqnarray}
Combining this with eq. \eqref{21} yields the important identity
\begin{equation}\label{27}
(D-1)\frac{\rho'}{\rho}\frac{\sigma'}{\sigma}=4\Lambda\sigma^{-1}-2\frac{\sigma''}{\sigma}-2
\left(\frac{\sigma'}{\sigma}\right)^2.
\end{equation}
It indeed allows us to express $\rho'/\rho$ in terms of $\sigma$ and its derivatives (for $\sigma'\neq 0$)
\begin{equation}\label{28}
\frac{\rho'}{\rho}=\frac{1}{D-1}
\left\{\frac{4\Lambda}{\sigma'}-2\frac{\sigma''}{\sigma'}-2\frac{\sigma'}{\sigma}\right\},
\end{equation}
and therefore
\begin{eqnarray}\label{29}
\frac{\rho''}{\rho}&=&\frac{1}{(D-1)^2}\left\{\frac{4\Lambda}{\sigma'}-2
\frac{\sigma''}{\sigma'}-2\frac{\sigma'}{\sigma}\right\}^2\nonumber\\
&&-\frac{1}{D-1}\left\{\frac{4\Lambda\sigma''}{\sigma'^2}+\frac{2\sigma'''}{\sigma'}-2
\left(\frac{\sigma''}{\sigma'}\right)^2\right.\nonumber\\
&&\left.+2\frac{\sigma''}{\sigma}
-2\left(\frac{\sigma'}{\sigma}\right)^2\right\}.
\end{eqnarray}

Using the variable
\begin{equation}\label{30}
s=\ln\left(\frac{\sigma}{\sigma_0}\right)
\end{equation}
we can now write eq. \eqref{23} as a third order differential equation
\begin{eqnarray}\label{31}
&&s'''-\frac{D}{D-1}\frac{(s'')^2}{s'}-\frac{4}{D-1}s''s'-\frac{D+3}{D-1}(s')^3\nonumber\\
&&+\frac{2}{D-1}\frac{\Lambda}{\sigma_0}e^{-s}
\left\{(D+1)\frac{s''}{s'}+(D+3)s'\right\}\nonumber\\
&&-\frac{4}{D-1}\frac{\Lambda^2}{\sigma^2_0}\frac{e^{-2s}}{s'}
=\frac{E^2}{\sigma^4_0}e^{-4s}s'\rho^{-(D-1)}.
\end{eqnarray}
For $E=0$ the scalar field plays no role and eq. \eqref{31} describes possible solutions to higher dimensional Einstein gravity. In this case no explicite $\rho$-dependence is left in eq. \eqref{31}. For $\Lambda=0$, one has a second order differential equation for $U=s'$, which has been discussed extensively in \cite{RDW}, \cite{CWCON}, and, for the special case $D=2$, in \cite{RSW}. This equation typically has solutions with singularities. 

In terms of the variables $s$ and 
\begin{equation}\label{31A}
v=(D-1)\ln\frac{\rho}{\rho_0}+4\ln\frac{\sigma}{\sigma_0}
\end{equation}
eqs. \eqref{23} and \eqref{27} read
\begin{eqnarray}\label{31B}
&&s''+\frac12 v's'-2\frac{\Lambda}{\sigma_0}e^{-s}=0,\\
&&v''+\frac{1}{2(D-1)}\big[v'^2-8v's'+4(D+3)s'^2\big]\label{31C}\\
&&+2E^2\sigma^{-4}_0\rho^{-(D-1)}_0e^{-v}=0.\nonumber
\end{eqnarray}
Eqs. \eqref{31B} and \eqref{31C} are equivalent to the two linear combinations of eqs. \eqref{19}-\eqref{21} that are independent of $C$. As long as $\sigma'\neq 0$ (or $s'\neq 0)$, they are also equivalent to the coupled system of equations \eqref{31} and \eqref{28}. In the special situation $E=0$, the latter equations decouple in the sense that one can solve first eq. \eqref{31} independently of an explicite computation of $\rho$, and subsequently determine $\rho$ from eq. \eqref{28}. The use of eqs. \eqref{31B}, \eqref{31C} has the advantage that the special solutions with flat four-dimensional space $(\Lambda=0)$ or constant scalar field $(E=0)$ are particularly apparent. An extensive discussion of the solutions for $\Lambda=E=0$ can be found in ref. \cite{CWCON}. 

\medskip\noindent
{\bf 3. Initial conditions and integration constants}

We recall, however, that the initial conditions for the numerical solution are constrained by eq. \eqref{21}. It is at this place that the geometry of the $D-1$ dimensional subspace, i.e. the constant $C$, enters. We  may choose the initial condition for the solution of eqs. \eqref{31B}, \eqref{31C} at some point $z_0$. Instead of the two initial values $s(z_0)$ and $v(z_0)$ we may consider $\sigma_0$ and $\rho_0$ as the corresponding integration constants. This allows us to use for all solutions a fixed prescription of the additive constants in $s$ and $v$, as, for example, $s(z_0)=v(z_0)=0$. The dependence on $\sigma_0$ and $\rho_0$ can be partly absorbed in the definitions
\begin{equation}\label{31D}
\tilde\Lambda=\frac{\Lambda}{\sigma_0},~\tilde E^2=E^2\sigma^{-4}_0\rho_0^{-(D-1)}~,~
\tilde C=\frac{(D-1)C}{\rho_0}.
\end{equation}
At this level we have three integration constants $\tilde\Lambda,\tilde E^2$ and $\tilde C$, while the overall normalization of $\sigma_0$ is irrelevant and plays no role after the redefinition of the integration constants. We remain with two free further integration constants, e.g. $s'(z_0)$ and $v'(z_0)$. Only one of them is independent. Indeed, eq. \eqref{21} yields the relation 
\begin{eqnarray}\label{31E}
&&\frac{D-2}{4}v'^2+2v's'-(D+3)s'^2\nonumber\\
&&=(D-1)\left[\tilde E^2e^{-v}+\tilde Ce^{\frac{4s-v}{D-1}}+4\tilde\Lambda e^{-s}\right].
\end{eqnarray}
Furthermore, we can often choose $z_0$ conveniently such that either $v'(z_0)$ or $s'(z_0)$ has a standard value, as $\pm 1$ or $0$. we are thus left with three continuous physical integration constants, namely $\tilde E,\tilde C$ and $\tilde\Lambda$. Due to translation invariance in $z$, we also may choose for $z_0$ a standard value, as $z_0=0$ or $z_0=1$. 

\section{Structure of singularities}
\label{structure}

The generic solutions to the field equations will have one or two singularities. This is reminiscent of Einstein gravity in four dimensions in absence of matter, where the nontrivial static radially symmetric solutions all show a singularity - the black hole. It is therefore useful to classify the singularities. This can be done with the help of eq. \eqref{27}, which relates a possible singular behavior of $\sigma$ with the one for $\rho$.

\newpage\noindent
{\bf 1. Powerlike singularities}

We first look at possible singularities with a powerlike behavior
\begin{equation}\label{32}
\sigma=\sigma_0z^{-\eta}~,~\rho=\rho_0z^\gamma,
\end{equation}
where eq. \eqref{27} implies, for $\eta>-2,\eta\neq 0,$
\begin{equation}\label{33}
\gamma=\frac{2+4\eta}{D-1}.
\end{equation}
Indeed, for $\eta>-2$ we can indeed neglect the terms $\sim \Lambda$ in the vicinity of the singularity. Similarly, for $\gamma<2$ (or $\eta<\frac D2-1$) we can neglect the terms $\sim C$. For the ``internal volume factor'' the relation \eqref{33} implies
\begin{equation}\label{49A}
\rho^{\frac{D-1}{2}}\sigma^2\sim z,
\end{equation}
such that the volume integration near the singularity remains finite. 

On the other hand, one finds
\begin{equation}\label{34}
\delta'^2=\tilde E^2z^2,\quad\tilde E^2=E^2\sigma^{-4}_0\rho^{-(D-1)}_0,
\end{equation}
such that the terms $\sim E^2$ influence the behavior near the singularity. We may determine $\eta$ from eq. \eqref{21}
\begin{eqnarray}\label{35}
\tilde E^2&=&3\eta^2-2(D-1)\eta\gamma+\frac14(D-1)(D-2)\gamma^2\nonumber\\
&=&\frac{1}{D-1}\big\{D-2-4\eta-(D+3)\eta^2\big\},
\end{eqnarray}
or
\begin{equation}\label{36}
\eta=\frac{1}{D+3}
\left(-2\pm\sqrt{(D-1)\big[D+2-\tilde E^2(D+3)\big]}\right).
\end{equation}
This leads us to the interesting observation that the strength of the singularity, as expressed by the ''singular exponent'' $\eta$, depends on the integration constant $\tilde E$. This differs from earlier examples \cite{RSW}, \cite{RDW}, \cite{CWCC}, \cite{CWCON} where $\eta$ was found to be determined only by $D$.

We note that for $\gamma<2$, where the terms $\sim C$ can be neglected near the singularity, the singular solution \eqref{32}, \eqref{33}, \eqref{36} automatically obeys eq. \eqref{26} and therefore solves the complete system \eqref{19}-\eqref{21}. This is not guaranteed for $\gamma=2$, where eq. \eqref{26} (or eq. \eqref{19}) yields an additional constraint for $\tilde C$. 

The maximal and minimal values of $\eta$ are reached for $\tilde E\to 0$
\begin{eqnarray}\label{36A}
&&\eta_+=\left[\sqrt{(D-1)(D+2)}-2\right]/(D+3),\nonumber\\
&&\eta_-=-\left[\sqrt{(D-1)(D+2)}+2\right]/(D+3),
\end{eqnarray}
with $\eta_\pm(D)$ taking the values $\eta_+(2)=0,~\eta_+(3)=(\sqrt{10}-2)/6,~\eta_+(\infty)=1,~\eta_-(2)=-4/5,~\eta_-(3)=-(\sqrt{10}+2)/6,\eta_-(\infty)=-1$. For all finite $D$ the allowed values of $\eta_\pm$ are in the range $-1<\eta_\pm<1$. For $\tilde E^2>0$ the two solutions obey $\eta_-<\eta_{1,2}<\eta_+$ and approach each other as $\tilde E^2$ increases. They coincide for $\tilde E^2_c=(D+2)/(D+3)$, with 
\begin{equation}\label{36B}
\eta(\tilde E_c)=-\frac{2}{D+3}.
\end{equation}
No solutions with real $\eta$ exist for $\tilde E^2>\tilde E^2_c$. We finally note that for $D=1$ all solutions with a powerlike singularity have $\eta=-1/2,~\tilde E^2=3/4$. (The quantities $\gamma$ and $C$ have no meaning in this case.)

\medskip\noindent
{\bf 2. Special cases}

In principle, we could have modifications of eq. \eqref{36} for the special case
\begin{equation}\label{37}
\gamma=2~,~\eta=\frac{D}{2}-1.
\end{equation}
Now the terms $\sim C$ can no longer be neglected and one finds from eq. \eqref{21} the condition
\begin{equation}\label{38}
\tilde C+\frac14(D-2)(D+2)+\tilde E^2=0.
\end{equation}
Such solutions can exist for $C<0$, or for $D=2$ if $\tilde E=0$ ($C$ vanishes for $D=2$), or for $D=1$ for $\tilde E^2=\frac34$. The choice $C<0$ leads, however, to a contradiction with eq. \eqref{19} and must be excluded. The cases with $\tilde C=0$ are contained in eqs. \eqref{33}, \eqref{36}, which therefore describe  all possible singular behaviors. For singularities obeying eq. \eqref{36A}, we observe that $\gamma<2$ holds for all $D>2$, while $\gamma=2$ is reached for $D=2$ and $\eta=\eta_+~(\tilde E=0)$. In conclusion, the term $\sim C$ can always be neglected close to the singularity, if $\eta\neq 0,\eta>-2$. 

As particular cases we may investigate singularities consistent with eq. \eqref{27}, which show singular behavior only in the function $\rho$, while $\sigma$ is regular. In the vicinity of the singularity at $z=0$ we may study the ansatz
\begin{equation}\label{39}
\rho=\rho_0z^\gamma,~\sigma=\sigma_0\left(1+\frac12\kappa z^2\right),
\end{equation}
for which eq. \eqref{27} implies for $z\to 0$
\begin{equation}\label{40}
\big[(D-1)\gamma+2\big]\kappa=4\Lambda\sigma^{-1}_0. 
\end{equation}
For $\gamma<2$, the leading singularities in eq. \eqref{21} cancel for
\begin{equation}\label{41}
\gamma=\frac{2}{D-1},~\tilde E^2=\frac{D-2}{D-1},~\kappa=\Lambda\sigma^{-1}_0.
\end{equation}
Similarly, solutions with singular $\sigma$ and regular $\rho$ are possible for 
\begin{equation}\label{42}
\sigma=\sigma_0z^{1/2},~\rho=\rho_0,~\tilde E^2=\frac34.
\end{equation}
The singularities \eqref{41}, \eqref{42} can be regarded as special cases of eqs. \eqref{32}, \eqref{33},  \eqref{36}.

Finally, an interesting class of possible solutions are characterized by the limiting behavior for $z\to 0$
\begin{equation}\label{43}
\rho=\rho_0z^2,~\sigma=\sigma_0.
\end{equation}
For $C=D-2,~E=0,~\Lambda=0,~\rho_0=1$ this describes flat space with a constant scalar field $\delta$ - an obvious solution of the field equations. (A flat torus, $C=E=\Lambda=0,~\rho=\rho_0,~\sigma=\sigma_0$, is also a solution.) More generally, for $\rho_0=C/(D-2)$ the point $z=0$ describes a regular fixed point of the rotational isometries acting on the $D-1$ dimensional subspace parameterized by $\bar y^\alpha$. (For the sphere $S^{D-1}$, with $C=D-2$, the isometries form the group $SO(D)$ and the rotational isometries are $SO(D-1)$.) For $D=2$ one has $C=0$ and $\rho_0\neq 1$ describes a canonical singularity \cite{CWCC}.

For a behavior of the type \eqref{43} we find a constraint from eq. \eqref{27}
\begin{equation}\label{44}
\frac{(D-1)(D-2)}{z^2}-\frac{\tilde C}{z^2}=\tilde E^2z^{-2(D-1)}.
\end{equation}
For $D>2$ such solutions are only possible for $\tilde E=0$ and $C/\rho_0=D-2$. In particular, all regular solutions of the type \eqref{43} have a constant scalar field. For $D=2$ the l.h.s. of eq. \eqref{44} vanishes, implying again $\tilde E=0$ and a constant scalar field. The solutions with a constant scalar field and a behavior \eqref{43} have been investigated in ref. \cite{RDW}. 

\medskip\noindent
{\bf 3. Exact power solutions}

For the special case $\Lambda=0,~C=0$ the powerlike solutions \eqref{32} are exact. They are actually the most general solutions. Indeed, we can combine eqs. \eqref{31C} and \eqref{31E} to 
\begin{equation}\label{S1}
v''+\frac12 v'^2=0.
\end{equation}
The general solution
\begin{equation}\label{S2}
v'=\frac 2z~,~v=2\ln z+v_0
\end{equation}
has used a first integration constant in order to have the singularity at $z=0$. Eq. \eqref{31B} takes the form
\begin{equation}\label{S3}
s''+\frac1z s'=0
\end{equation}
with general solution
\begin{equation}\label{S4}
s'=-\frac\eta z~,~s=-\eta\ln z+s_0.
\end{equation}
We can set $v_0=s_0=0$ since these integration constants can be absorbed by $\sigma_0$ and $\rho_0$. This yields
\begin{equation}\label{S5}
\sigma=\sigma_0z^{-\eta}~,~\rho=\rho_0z^\gamma~,~\gamma=\frac{2+4\eta}{D-1}
\end{equation}
as the most general solution. Insertion into eq. \eqref{31E} finally determines $\eta$ according to eq. \eqref{36}. No further constraint restricts $\rho_0$, which is therefore a free integration constant. 

\section{Warped solutions and local four-dimensional gravity}
\label{warpedsolutions}

In this section we show that all realistic solutions with a consistent four-dimensional gravity should  have two singularities if $E\neq 0$. For this purpose we have to classify the possible behavior for $z\to\pm\infty$, and to define the criterion for a finite effective four-dimensional gravitational constant. More generally, we discuss properties of the solutions which lead to an effective four-dimensional gravity at long distances. 

\medskip\noindent
{\bf 1. Number of singularities and types of warped 

\hspace{0.2cm}solutions}

Solutions with no or one singularity could approach finite nonzero values of $\sigma(z)$ and $\rho(z)$ as $z\to\infty$ or $z\to-\infty$. Eq. \eqref{27} tells us that this is only possible for $\Lambda=0$, while the combination of eqs. \eqref{19} and \eqref{21} yields $C=0,~E=0$, such that this option cannot be realized for $E\neq 0$. We conclude that all solutions with a non-constant scalar field must either have two singularities that we may locate at $z=0$ and $z=\bar z>0$ without loss of generality, or else at least one of the functions $\rho(z)$ or $\sigma(z)$ must approach zero or diverge as $z\to\infty$ or $z\to-\infty$.

For a discussion of the possible types of solutions for $z\to\pm\infty$ we may again assume a powerlike behavior \eqref{32}. Eq. \eqref{27} admits three types of solutions. For the first, $\sigma$ goes to a constant, vanishes or increases less fast than $z^2~(\eta>-2)$. This is only possible for $\Lambda=0$. The exponent $\gamma$ obeys eq. \eqref{33} and therefore $\gamma>-6/(D-1)$. For the second type, $\sigma$ increases $\sim z^2~(\eta=-2)$. This requires $\gamma=2(\tilde\Lambda-3)/(D-1)$. The third type of solutions has the leading behavior \eqref{33} with $\eta<-2,~\gamma=(2+4\eta)/(D-1)<-6/(D-1)$, and the term $\sim\Lambda$ is subdominant for the asymptotic behavior. The case $\eta=-2$ is the boundary between the first and third type of solutions. It is remarkable that for all solutions with an asymptotic behavior $\eta>-2$ the integration constant $\Lambda$ must vanish. On the other hand, eq. \eqref{21} can be obeyed for $C\neq 0$ only if $\gamma\geq 2$. Solutions of the third type could therefore exist only for $C=0$. In this case, however, eq. \eqref{36} applies and therefore $\eta>-1$. We conclude that the third type of solutions cannot be realized. For the boundary case we consider first $\gamma>2$ or $C=0$. Then the term $\sim C$ in eq. \eqref{21} can be neglected and we find for $\eta=-2$ the relations
\begin{eqnarray}\label{45}
\eta=-2,\tilde E^2&=&\frac14 (D-1)(D-2)\gamma^2+2(D-1)\gamma,\nonumber\\
\tilde\Lambda&=&\frac12(D-1)\gamma+3.
\end{eqnarray}
For the special case $\gamma=2$, eq. \eqref{21} relates $\tilde E^2$ to $\tilde C$
\begin{equation}\label{46}
\eta=-2,~\gamma=2,~\tilde E^2=(D-1)(D+2)-\tilde C,~\tilde\Lambda=D+2.
\end{equation}

For the first type of solutions with $\tilde \Lambda=0$, eq. \eqref{36} holds if $C=0$ or $\gamma>2$. Then the maximal value for $\gamma$ is $6/(D-1)$, excluding solutions with $\gamma>2$ for $D>3$. For $D=3$ the condition $\eta>1/2$, required for $\gamma>2$, is not compatible with eq. \eqref{36}. We conclude that all possible solutions of the first type $(\eta>-2)$ must have $\gamma=2$ or $C=0$. The solutions with $\gamma=2$ obey
\begin{equation}\label{47}
\eta=\frac D2-1~,~\gamma=2~,~\tilde E^2=-\frac14(D-2)(D+2)-\tilde C~,~\Lambda=0,
\end{equation}
while for $C=0$ one has a range of critical exponents related by eqs. \eqref{33} and \eqref{36} to $\tilde E^2$, 
\begin{equation}\label{48}
\eta>-2~,~\gamma>-\frac{6}{D-1}~,~C=0~,~\Lambda=0.
\end{equation}
Thus for $D>2$ all solutions of the first type need $C\leq 0$. At this stage, the possible asymptotic behaviors for $z\to\pm\infty$ are given by eqs. \eqref{45}-\eqref{48}. 

\medskip\noindent
{\bf 2. Requirement of consistent four-dimensional 

\hspace{0.2cm}gravity}

We next restrict our general discussion to solutions that lead to an acceptable four-dimensional gravity. After dimensional reduction the effective Planck mass $M=(8\pi G_N)^{-1/2}$ obeys
\begin{equation}\label{49}
M^2=M^{D+2}_dL^D\int_z\sigma\rho^{\frac{D-1}{2}}.
\end{equation}
Here we define $\int_z$ such that it includes a factor $V_{D-1}$ corresponding to the volume of the $D-1$ dimensional subspace parameterized by the coordinates $\bar y^\alpha$. For a powerlike behavior \eqref{32} near a singularity at $z=0$, or for $z\to \infty$, eq. \eqref{33} implies
\begin{equation}\label{50}
\sigma\rho^{\frac{D-1}{2}}\sim z^{-\eta+\frac{D-1}{2}\gamma}\sim z^{1+\eta}.
\end{equation}
Finiteness of the gravitational constant requires near a singularity $\eta>-2$, while for $z\to\infty$ one needs $\eta<-2$. If these conditions are not obeyed, an effective four-dimensional description of gravity at long distances is not possible. Dimensional reduction is not meaningful in this case. As we have seen above, asymptotic solutions for $z\to\infty$ with $\eta<-2$ are not compatible with our system of equations. In consequence, the only realistic solutions admitting four-dimensional gravity must have two singularities, one located at $z=0$, the other at $z=\bar z$. (For $E=0$, one of the singularities can be replaced by the regular behavior \eqref{43}.) 

The structure of the solutions near $z=\bar z$ is the same as for $z=0$,
\begin{equation}\label{51}
\sigma=\bar\sigma(\bar z-z)^{-\bar\eta},~\rho=\bar \rho(\bar z-z)^\gamma.
\end{equation}
However, the exponents $\eta$ and $\gamma$ may differ. This difference can arise from the different sign characterizing the two solutions \eqref{36}. Also the constants $\tilde E^2,~\tilde C,~\tilde \Lambda$ are now replaced by $\bar E^2, \bar C, \bar \Lambda$, which are defined similar to eq. \eqref{31D}, but with $\bar\sigma,\bar\rho$ replacing $\sigma_0,\rho_0$. For all solutions of eq. \eqref{36} one has $\eta>-1$, such that $M^2$ in eq. \eqref{49} is well defined. As we have discussed before, the general solution has three continuous integration constants which may be chosen as the constants $\tilde E^2,~\tilde C,~\tilde \Lambda$ in the vicinity of the singularity at $z=0$, with $\rho_0$ and $\sigma_0$ defined by eq. \eqref{32}. A given local solution in the vicinity of $z=0$ has then to be extended to larger $z$ until a new singularity is reached. 

\newpage
\section{Geometry near singularities}
\label{geometry}

\medskip\noindent
{\bf 1. Cusps, cones and infinitons}

Let us consider the geometry of the solutions with singularities somewhat closer. We first note that, in general, the singularities are true singularities, and not coordinate singularities. For $E\neq 0$ the curvature scalar diverges at the singularities,
\begin{eqnarray}\label{x1}
\tilde R(z\to 0)&&\to~E^2L^{-2}\sigma^{-4}\rho^{-(D-1)}\sim E^2 z^{4\eta-(D-1)\gamma}\nonumber\\
&&=E^2z^{-2},
\end{eqnarray}
and similar for $z\to\bar z$. This divergence is not integrable, since the volume element $\sim\int dz\sigma^2\rho^{\frac{D-1}{2}}\sim\int dzz$ leads to logarithmic singularities in $\int_z\tilde g^{1/2}\tilde R$. The trace of the energy momentum tensor shows a similar divergence, which can be considered as the source for the singularity in $\tilde R$. For the action, the two singularities cancel such that $\Gamma=0$. We note that $\delta(z)$ diverges logarithmically at the singularities. With
\begin{equation}\label{x2}
\delta'(z\to 0)\to E\sigma^{-2}\rho^{-\frac{D-1}{2}}\sim Ez^{-1}
\end{equation}
one finds 
\begin{equation}\label{x3}
\delta(z\to 0)\sim E\ln z~,~
\delta(z\to\bar z)\sim -E\ln (\bar z-z).
\end{equation}
Since $\sigma$ and $\rho$ are both positive, $\delta$ increases (decreases) monotonically for $E>0~(E<0)$. The action is symmetric with respect to a reflection $\delta\to-\delta$. Witout loss of generality we can therefore choose $E>0$. Then $\delta(z)$ approaches $-\infty$ for $z\to 0$, and $+\infty$ for $z\to\bar z$. 

Consider next the geometry of the $D$-dimensional internal space, i.e. the hypersurfaces with fixed $x^\mu, dx^\mu=0$. This geometry is described by $\rho(z)$ $\bar g_{\bar\alpha\bar\beta}(\bar y)$. As an example, $\bar g_{\bar\alpha\bar\beta}(\bar y)$ may be the metric of the $(D-1)$-dimensional unit sphere, with $C=D-2$. For $\gamma>0$ one finds a pointlike singularity - the ``radius'' of the $(D-1)$-dimensional subspace shrinks to zero as the singularity is approached. Only for $\rho=z^2$ the point $z=0$ could be included in the manifold, similar to the pole in the $D$-dimensional sphere. For $\gamma<2$ the internal curvature $R^{(D)}$ diverges as the singularity is approached. 
\begin{equation}\label{x4}
R^{(D)}=(D-1)\gamma(1-\frac D4\gamma)z^{-2}.
\end{equation}
For the range $0<\gamma<2$ we conclude
\begin{equation}\label{x5}
R^{(D)}(z\to 0)\to\left\{
\begin{array}{ccc}
\infty&for&\gamma<4/D\\
-\infty&for&\gamma>4/D\\
0&for&\gamma=4/D
\end{array}
\right..
\end{equation}
(Eqs. \eqref{x4}, \eqref{x5} hold for $D=2$ for all $\gamma$.) For $R^{(D)}\to-\infty~(\gamma>4/D)$ the singularity is a cusp, and for $R^{(D)}\to 0$ a generalized cone. (For $D=2$ the singularities $\rho\sim z^\gamma,~\gamma<2$ or $\rho=\alpha z^2,~\alpha>1$, cannot be embedded in flat three dimensional space.) The singularities with $\gamma>0$ can be associated with a generalized ``brane'' \cite{ARS}, \cite{7A}, \cite{PR} sitting at $z=0$. 

The brane or pointlike singularities correspond to $\eta>-1/2$. Depending on $\tilde E$, the solution of eq. \eqref{36A} with the plus sign corresponds to the range
\begin{equation}\label{x6}
\frac{2}{D+3}\leq\gamma\leq\frac{2}{D+3}
\left(1+2\sqrt{\frac{D+2}{D-1}}\right),
\end{equation}
such that the singularity is powerlike and may have positive or negative curvature. Cusp singularities with $\gamma>4/D$ are realized close to the upper bound $(\tilde E^2\to 0)$ for all $D>2$, whereas the lower bound corresponds to positive $R^{(D)}$. Also the solutions with a negative sign in eq. \eqref{36A} can have a pointlike singularity with $R^{(D)}\to\infty$ for sufficiently large $\tilde E^2$.

For $\eta\to\eta_-~(\tilde E^2\to 0)$, however, the exponent $\gamma$ becomes negative, with
\begin{equation}\label{x7}
\gamma_-=\frac{2}{D+3}\left(1-2\sqrt{\frac{D+2}{D-1}}\right).
\end{equation}
For $\gamma<0$, the singularity is of a different type. Close to the singularity, the ``radius'' of the $(D-1)$-dimensional subspace now diverges. The curvature $R^{(D)}$ also diverges to negative infinity. Singularities of this type may be called ``infinitons''. Nevertheless, the volume of internal space remains finite as long as $\gamma>-2/(D-1)$  This is always realized for $\gamma\geq\gamma_-$.

\medskip\noindent
{\bf 2. Warped branes and zerowarps}

Another important aspect of the singularities concerns the warp factor $\sigma$. Let us consider the five-dimensional space spanned by the coordinates $x^\mu$ and $z$. Apart from the signature, this is the analogue of $D$-dimensional internal space, with $\rho$ replaced by $\sigma$. Indeed, for $D=5$ our equations are invariant under the exchange $\sigma\leftrightarrow\rho,~\tilde C\leftrightarrow 4\tilde \Lambda$, or $\gamma\leftrightarrow-\eta$. For $\eta<0$, one has a pointlike singularity in five-dimensional space. As the singularity is approached for $z\to 0$, the five dimensional distance between two arbitrary points in four-dimensional space-time shrinks to zero. Even the locations of the most remote galaxies have an arbitrarily small five dimensional distance from each other. The singularities with $\eta<0$, where the warp factor $\sigma$ vanishes at the singularity, are called ``zerowarps''. 
For $\eta>0$, in contrast, the five dimensional distance between two locations $x^\mu$ and $x'^\mu$ diverges to infinity as $z\to 0$. For a fixed four-dimensional volume the volume of the five dimensional space remains finite as long as $\eta<1/2$. This is not automatically obeyed. For $D\geq 7$, an infinite volume of five-dimensional space becomes possible $(\eta>1/2)$, even though the four-dimensional volume is finite.

We conclude that the singularities either correspond to warped branes $(\gamma>0)$ or to zerowarps $(\eta<0)$. Singularities where a brane is simultaneously a zerowarp are possible only for $\tilde E^2$ exceeding a minimum value, not for $\tilde E=0$. Zerowarp solutions have first been discussed in \cite{RSW}, and warped brane solutions in \cite{7A}. An intensive discussion of warped branes with codimension one has stated with \cite{RS}, and codimension two branes are investigated more recently in \cite{GS,CS,9A,SCW}. We discuss warped branes in five-dimensional gravity with a dilaton (codimension one) in appendix A. Warped branes with codimension $\geq 2$, embedded in Ricci-flat higher dimensional space, are investigated in detail in \cite{CWCON}. 

\section{Global properties of solutions}
\label{global}

\medskip\noindent
{\bf 1. Monotonic behavior}

Let us next turn to the global properties of the solutions. This concerns, in particular, the question which type of singularities can be connected by a smooth solution for $0<z<\bar z$ (or $0< z<\infty$). We first consider the warp factor and ask if a $\sigma$ is monotonic or if a turning point (maximum or minimum of $\sigma(z)$) can occur. For a turning point with $\sigma'(z_t)=0$, eq. \eqref{27} implies $\sigma''(z_t)=2\Lambda$. For $\Lambda>0$ a minimum of $\sigma$ at $z_t$ is possible, and for $\Lambda<0$ a maximum may occur. For $\Lambda=0$ one finds $\sigma''(z_t)=0$ and higher derivatives of $\sigma$ at $z_t$ are needed for a characterization of a possible turning point. For $\Lambda=0$, one actually finds that $\sigma'(z_t)=0$ implies that all derivatives of $\sigma$ vanish at $z_t$. This follows by subsequent differentiations of eq. \eqref{27}. For $\Lambda=0$, we conclude that either $\sigma$ is monotonic or constant. For a monotonic $\sigma$ there must then be at least one singularity (or regular point) with $\eta\leq 0$, such that $\sigma$ vanishes or approaches a constant. On the other hand, either $\sigma$ diverges at a second singularity of for $z\to\pm\infty$, or it approaches a constant.

A similar discussion is possible for the potential turning points of $\rho$ at $z_\rho,~\rho'(z_\rho)=0$. A linear combination of eqs. \eqref{19} and \eqref{20} yields
\begin{eqnarray}\label{x8}
&&\Lambda\sigma^{-1}-\frac{D+5}{4}C\rho^{-1}-\frac34\left(\frac{\sigma'}{\sigma}\right)^2+\frac32\frac{\rho'}{\rho}
\frac{\sigma'}{\sigma}\nonumber\\
&&+\frac{D+2}{4}\frac{\rho''}{\rho}+\frac{D^2+D-14}{16}\left(\frac{\rho'}{\rho}\right)^2=-\frac14\delta'^2,
\end{eqnarray}
and, in combination with eq. \eqref{21}, we obtain
\begin{equation}\label{x9}
\frac{\rho'}{\rho}\frac{\sigma'}{\sigma}=C\rho^{-1}-\frac12\frac{\rho''}{\rho}-
\frac{D-3}{4}\left(\frac{\rho'}{\rho}\right)^2.
\end{equation}
For a possible turning point this yields
\begin{equation}\label{x10}
\rho''(z_\rho)=2C. 
\end{equation}
Similar to the discussion above, we conclude that for $C=0$ the function $\rho(z)$ is either monotonic or constant. For $C>0$ it may have a minimum, but no maximum. In particular, for $C\geq 0$ the presence of two pointlike singularities with $\gamma>0$ is excluded. 

\medskip\noindent
{\bf 2. Solutions with $\Lambda = 0$}

In the following we will consider $\Lambda=0$. For $C=0$ we have found that the powerlike solutions \eqref{32}, with $\eta$ and $\gamma$ obeying eqs. \eqref{33}, \eqref{36}, are exact solutions for the whole range of $z$. These exact solutions extend from $z=0$ to $z\to\infty$ and have therefore no second singularity at some finite $\bar z$. They correspond to the third type of asymptotic solutions for $z\to\infty$, eq. \eqref{48}, and do not admit dimensional reduction with finite $M^2$, eq. \eqref{49}. For $C\neq 0$, the powerlike solutions are approached for $\tilde C\to 0$. Typically the powerlike solutions are dividing lines between solutions with two singularities (for $\tilde C>0$)  and solutions with a regular behavior for $z\to\infty$ (for $\tilde C<0$). For $\Lambda=0$ one has still two free integration constants left, while the singular behavior depends only on one integration constant, $\tilde E^2$ or $\eta$. The second integration constant does not influence the behavior at the pole for $z\to 0$, but it decides on which side of the deviding line a given solution is located. For $C\neq 0$ we may use $\tilde C$ as the second integration constant. 

\medskip\noindent
{\bf 3. Classification of solutions}

We may classify the possible solutions according to the type of the singularity for $z\to 0$. For type $(A)$, the solution corresponds to the plus sign in eq. \eqref{36A} with $\eta>0$, i.e.
\begin{equation}\label{y1}
(A):\quad \eta>0~,~\gamma>0~,~\tilde E^2<\frac{D-2}{D-1}.
\end{equation}
Solutions of type $(B)$ again have the plus sign in eq. \eqref{36A} but $\eta<0$
\begin{equation}\label{y2}
(B):\quad \eta<0~,~\gamma>0~,~
\frac{D-1}{D+2}<\tilde E^2\leq\frac{D+2}{D+3}.
\end{equation}
Similarly, the type $(C)$ characterizes solutions with the minus sign in eq. \eqref{36A}, with $\gamma>0$
\begin{equation}\label{y3}
(C):\quad \eta<0~,~\gamma>0~,~\frac34<\tilde E^2\leq\frac{D+2}{D+3}.
\end{equation}
The singularities $(A)-(C)$ describe warped branes. The solutions $(B)$ and $(C)$ are simultaneously zerowarps. Finally, the type $(D)$ denotes zerowarp solutions where both $\eta$ and $\gamma$ are negative, with a minus sign in eq. \eqref{36A},
\begin{equation}\label{y4}
(D):\quad \eta<0~,~\gamma<0~,~\tilde E^2<\frac34.
\end{equation}
For $\tilde E^2=(D+2)/(D+3)$ the types $(B)$ and $(C)$ coincide. The boundary between $(C)$ and $(D)$ has $\tilde E^2=3/4,~\gamma=0,~\eta=-\frac12$, and the boundary between $(A)$ and $(B)$ obeys $\tilde E^2=(D-2)/(D-1), ~\eta=0,~\gamma=2/(D-1)$.

\medskip\noindent
{\bf 4. Solutions with two singularities}

Let us now investigate the possibilities for two singularities. The monotonic behavior of $\sigma(z)$ tells us that  a solution of type $(A)$ can only be connected to a second singularity at $z=\bar z$ which is of type $(B),~(C)$ or $(D)$. Similarly, singularities at $z=0$ of the types $(B), ~(C)$ or $(D)$ require a type $(A)$ singularity at $z=\bar z$. Thus one of the singularities is always a pointlike singularity of type $(A)$. Since the role of the two singularities is interchangeable, we may take for $z\to 0$ the solution of type $(A)$. 

For $C>0$ the second singularity at $\bar z$ cannot be pointlike. The only remaining possibility is a singularity of type $(D)$. (Only for $C<0$ the type $(B)$ or $(C)$ singularities at $\bar z$ are also conceivable.) Let us concentrate on $C>0$. The solution starts for $z\to 0$ with $\rho\to 0,~\sigma\to\infty$ according to
\begin{eqnarray}\label{y5}
\rho&=&\rho_0 z^\gamma~,~\frac{2}{D-1}<\gamma<\gamma_+\nonumber\\
\sigma&=&\sigma_0z^{-\eta}~,~0<\eta<\eta_+,
\end{eqnarray}
with $\eta_+$ given by eq. \eqref{36B}. Correspondingly, the variables $s$ and $v$ (eqs. \eqref{30}, \eqref{31A}) obey
\begin{equation}\label{y6}
\hat s=-\eta\ln z~,~\hat v=2\ln z.
\end{equation}
Close to the singularity we can linearize eqs. \eqref{31B}, \eqref{31C}, \eqref{32} with
\begin{equation}\label{y7}
s(z)=\hat s(z)+\delta_s(z)~,~v(z)=\hat v(z)+\delta_v(z),
\end{equation}
such that
\begin{eqnarray}\label{Y8}
&&z^2\delta''_s+z\delta'_s-\frac{\eta z}{2}\delta'_v=0,\nonumber\\
&&z^2\delta''_v+\frac{2z}{(D-1)}
\Big[(1+2\eta)\delta'_v-\big(4+2(D+3)\eta\big)\delta'_s\Big]\nonumber\\
&&-2\tilde E^2\delta_v=0,\nonumber\\
&&\frac{2z}{D-1}\Big[\left(1+2\eta-\frac{D}{2}\right)\delta'_v-\big(4+2(D+3)\eta\big)\delta'_s\Big]\nonumber\\
&&-2\tilde E^2\delta_v+2\tilde C z^{2-\gamma}=0.
\end{eqnarray}
Combining the last two equations yields
\begin{equation}\label{Y9}
z^2\delta''_v+\frac{Dz}{D-1}\delta'_v-2\tilde Cz^{2-\gamma}=0.
\end{equation}
A solution is
\begin{eqnarray}\label{Y10}
&&\delta_v=\frac{2\tilde C}{(2-\gamma)\left(2-\gamma+\frac{1}{D-1}\right)}z^{2-\gamma},\nonumber\\
&&\delta_s=\frac{\eta\tilde C}{(2-\gamma)^2\left(2-\gamma+\frac{1}{D-1}\right)}z^{2-\gamma}
\end{eqnarray}

For $\Lambda=0$ and given $\tilde E^2$, let us consider the limit $\tilde C\to 0_+$. In this limit the location of the second singularity $\bar z$ will move to infinity. The solution stays very near the exact solution for a large range $0<z<z_c$, with $z_c$ of the order of $\bar z$ but somewhat smaller. In the region $z\approx z_c$ the power law solution \eqref{32}, \eqref{33}, \eqref{36} around $z=0$ is matched to a similar singular solution around $z=\bar z$. 

\section{Dimensional reduction}
\label{dimensional}

At this stage we have described the general quasistatic solutions of the field equations \eqref{2}, \eqref{2a} that are consistent with the ansatz \eqref{14}. They are described by several free integration constants. In particular, the cosmological constant $\Lambda$ turns out to be one of these integration constants. In the presence of singularities not all solutions of the higher dimensional field equations correspond to extrema of the action \cite{CWCON}. One possible way to understand this issue investigates the effective four-dimensional action.

\medskip\noindent
{\bf 1. Four-dimensional effective action}

The four-dimensional action will also be useful for a discussion of cosmological solutions. The quasistatic solutions discussed so far are possible candidates for the asymptotic state of the universe as $t\to\infty$, but cannot describe the cosmological evolution. In particular, the cosmological evolution will decide which one of the possible asymptotic states is approached. 

An investigation of the general time dependent solutions is rather involved. The functions $\sigma,\rho$ and $\delta$ become now functions of $z$ and $t$, and the most general ansatz consistent with internal symmetries, time and space translations, as well as rotations and reflections in space, contains several other functions of $z$ and $t$ \cite{CWS2}. In order to get a first approach, one may  study the cosmological solutions of an effective four-dimensional theory, obtained by dimensional reduction.

Dimensional reduction involves an ansatz for the relevant degrees that may be important for cosmology after some early initial period - say after inflation. Besides the four-dimensional metric $\tilde g^{(4)}_{\mu\nu}(\vec x,t)$, this typically comprises other fields with a small mass. We will include here the radion by promoting $L$ in eq. \eqref{14} to a field $L(\vec x,t)$. In addition, we consider the dilaton by the ansatz $\delta(x,y)=\delta^{(0)}(z)+\delta(\vec x,t)$. The reduced four-dimensional theory is found by inserting the ansatz in the higher dimensional action \eqref{10} and integrating over the internal coordinates $y^\alpha$, 
\begin{equation}\label{A1a}
\Gamma^{(4)}[\tilde g^{(4)}_{\mu\nu}, L,\delta]=\frac{M^{d-2}_d}{2}\int_y\tilde g^{1/2}\{-\tilde R+\partial^{\hat \mu}\delta
\partial_{\hat\mu}\delta\}.
\end{equation}
For this purpose we have to specify appropriate functions $\sigma^{(0)}(z),~\rho^{(0)}(z),~\bar g^{(0)}_{\bar\alpha\bar\beta}(\bar y),~\delta^{(0)}(z)$. We will choose them as one of our quasistatic solutions. This choice is not unique, and we will discuss this issue in more detail below.

With our ansatz for the metric we can decompose the higher dimensional curvature scalar (omitting terms that will lead to irrelevant total derivatives)
\begin{eqnarray}\label{A2}
\tilde R&=&\tilde R^{(4)}(x)\sigma^{-1}(z)-\tilde c\partial^\mu\ln L(x)\partial_\mu\ln L(x)\sigma^{-1}(z)\nonumber\\ 
&&+L^{-2}(x)\tilde R_D(z),
\end{eqnarray}
with
\begin{equation}\label{A3}
\tilde R_{(D)}(z)=R^{(D)}-\left(\frac{\sigma'}{\sigma}\right)^2-4\frac{\sigma''}{\sigma}-2(D-1)\frac{\rho'}{\rho}\frac{\sigma'}{\sigma}.
\end{equation}
Here 
\begin{equation}\label{62A}
R^{(D)}=\frac{(D-1)C}{\rho}-(D-1)\frac{\rho''}{\rho}-\frac14(D-1)(D-4)\left(\frac{\rho'}{\rho}\right)^2
\end{equation}
is the curvature scalar computed from the internal part of the metric \eqref{14} for $L=1$. It depends on $\rho,\rho'$ and $\rho''$. (We abbreviate here $\sigma,\rho$ instead of $\sigma^{(0)}(z),~\rho^{(0)}(z)$.) Inserting into the higher dimensional action yields
\begin{eqnarray}\label{A4}
&&\Gamma^{(4)}=\int_x(\tilde g^{(4)})^{1/2}\\
&&\left\{\frac12\chi^2[-\tilde R^{(4)}+\tilde c\partial^\mu\ln L~\partial_\mu\ln L+\partial^\mu\delta~\partial_\mu\delta]+V
\right\},\nonumber
\end{eqnarray}
where
\begin{equation}\label{A5}
\chi^2=M^2_d(M_d L)^D\int_z\sigma^{-1}\rho^{\frac{D-1}{2}},
\end{equation}
and
\begin{equation}\label{A6}
V=\frac12 M^4_d(M_d L)^{D-2}\int_z\sigma^2\rho^{\frac{D-1}{2}}
\{\delta'^2-\tilde R_D\}.
\end{equation}

\medskip\noindent
{\bf 2. Four-dimensional solutions and contradiction 

\hspace{0.2cm}for $\Lambda \neq 0$}

For any solutions of the higher dimensional field equations the higher dimensional action and therefore also $\Gamma^{(4)}$ must vanish. This follows directly from eq. \eqref{13}. We employ this observation for a computation of $V$ by noting that for $\partial_\mu\ln L=0,~\partial_\mu\delta=0,~\tilde R^{(4)}=4\Lambda L^{-2}$ we have indeed a solution of the higher dimensional field equations, implying
\begin{equation}\label{A7}
\delta'^2-\tilde R_D=4\Lambda\sigma^{-1}.
\end{equation}
(This can, of course, be verified by a direct computation for our quasistatic solutions.) We find
\begin{equation}\label{A8}
V=2\Lambda M^4_d(M_d L)^{D-2}\int_z\sigma\rho^{\frac{D-1}{2}}=2\Lambda\chi^2L^{-2}.
\end{equation}

For constant $L$ and $\delta$ one may be tempted to relate $V$ and the effective four-dimensional cosmological constant $\Lambda$ by eq. \eqref{A8}. Considering only the higher dimensional solutions of the field equation this seems possible for arbitrary $\Lambda$. The situation is more complicated, however. Solving the field equations for the effective four-dimensional theory with $L=\delta=const.$ results in 
\begin{equation}\label{A9}
\chi^2(\tilde R^{(4)}_{\mu\nu}-\frac12\tilde R^{(4)}\tilde g^{(4)}_{\mu\nu})=-V\tilde g^{(4)}_{\mu\nu}
\end{equation}
or
\begin{equation}\label{A10}
\chi^2\tilde R^{(4)}=4V.
\end{equation}
This implies $\tilde R^{(4)}=4V/\chi^2=8\Lambda L^{-2}$, leading to a contradiction with the higher dimensional solution $\bar R^{(4)}=4\Lambda L^{-2}$, unless $\Lambda=0$. This simple observation singles out the case of a vanishing cosmological constant $\Lambda=0$. Only in this case the higher dimensional and four-dimensional field equations lead to the same static solution. 

For higher dimensional solutions with integration constant $\Lambda=0$ the potential $V(L,\delta)$ vanishes identically. In this case $L=const.,~\delta=const.$ are indeed solutions of the four-dimensional field equations. Again, this does not hold for $\Lambda\neq 0$. Now we find a nonvanishing potential $V\sim L^{D-2}$. 

\medskip\noindent
{\bf 3. Four-dimensional Weyl scaling}

Since also the effective four-dimensional gravitational constant $\sim \chi^{-2}$ depends on the field $L$, a better understanding of the cosmological solutions of the field equations derived from the action \eqref{A4} can be gained after Weyl scaling. With
\begin{eqnarray}\label{A11}
\tilde g^{(4)}_{\mu\nu}&=&w^2_4 g_{\mu\nu}~,~w_4=M\chi^{-1},\\
\tilde R^{(4)}&=&w^{-2}_4(R-f_4\partial^\mu\ln w_4~\partial_\mu\ln w_4-g_4D^2\ln w^2_4),\nonumber
\end{eqnarray}
and $R$ the curvature scalar corresponding to $g_{\mu\nu}$, one has $(f_4=6)$
\begin{eqnarray}\label{A12}
\Gamma^{(4)}&=&\int_xg^{1/2}\left\{\frac12 M^2\Big[-R+f_4\partial^\mu \ln\chi~\partial_\mu\ln\chi\right.\nonumber\\
&&\left.+\tilde c\partial^\mu\ln L\partial_\mu\ln L+\partial^\mu\delta~\partial_\mu\delta\Big]+\frac{M^4}{\chi^4}V
\right\}\nonumber\\
&=&\int_xg^{1/2}\left\{\frac12 M^2\Big[-R+\left(6+\frac{4\tilde c}{D^2}\right)\partial^\mu\ln\chi\partial_\mu\ln\chi\right.\nonumber\\
&&\left.+\partial^\mu\delta~\partial_\mu\delta\Big]
+2\Lambda\hat c^{\frac 2D}M^4\left(\frac{\chi^2}{M^2_d}\right)^{-\frac{D+2}{D}}\right\},
\end{eqnarray}
where
\begin{equation}\label{A13}
\hat c=\int_z\sigma\rho^{\frac{D-1}{2}}.
\end{equation}

We may introduce canonically normalized fields for the dilaton, $\Delta=M\delta$, and radion
\begin{equation}\label{A14}
\tilde\varphi=\left(6+\frac{4\tilde c}{D^2}\right)^{1/2}M\ln\frac{\chi}{M_d}
\end{equation}
such that their kinetic terms have the standard form for real scalar fields. For $\Lambda\neq 0$ we find an effective potential for the radion, $U=(M^4/\chi^4)V$, 
\begin{eqnarray}\label{A15}
U&=&2\Lambda\hat c^{\frac2D}M^4\exp\left(-\tilde\alpha\frac{\tilde\varphi}{M}\right),\nonumber\\
\tilde\alpha&=&\frac{2(D+2)}{D}\left(6+\frac{4\tilde c}{D^2}\right)^{-1/2}.
\end{eqnarray}

\medskip\noindent
{\bf 4. Cosmological solutions}

We recognize the radion as a type of quintessence field that increases towards infinity as time goes on. For $\Lambda\neq 0$, the effective four-dimensional theory describes cosmologies with an asymptotically vanishing dark energy, and not a de Sitter space with a cosmological constant different from zero. Asymptotically, the curvature scalar in the Einstein frame $R$ goes to zero. In the absence of matter the asymptotic solution is \cite{CWQ}, \cite{CWAA}
\begin{equation}\label{A16}
\frac{\tilde\varphi}{M}=\frac {2}{\tilde\alpha}\ln t+\tilde\varphi_0.
\end{equation}

If we expand around a higher dimensional solution $\big(\sigma^{(0)}(z)~,~\rho^{(0)}(z)~,~\tilde g^{(0)}_{\alpha\beta}(\bar y)~,~\delta^{(0)}(z)\big)$ with integration constant $\Lambda\neq 0$ we conclude that this solution is not a solution of the field equations for the effective four-dimensional theory. The four-dimensional solution leads to asymptotically flat space $(R\to 0)$, but increasing $\chi$ and therefore $L$. This is in contradiction to the higher dimensional solution with constant $\tilde R^{(4)}$ and $L$. For an extremum of $\Gamma$ the four-dimensional field equations must be obeyed, however. We infer that higher dimensional solutions with integration constant $\Lambda\neq 0$ cannot be extrema of the effective action. In the presence of singularities the extrema of the action have to obey additional constraints beyond the higher dimensional field equations \cite{CWCON}. We will address this issue in the next section. We will see that these constraints allow quasistatic solutions only for $\Lambda=0$.

\section{Potential for integration constants}
\label{potentialfor}

The action $\Gamma$ is a functional of arbitrary metrics $\tilde g_{\hat\mu\hat\nu}(x,y)$ and scalar fields $\delta(x,y)$. The extremum condition $\delta \Gamma=0$ singles out the allowed solutions. If we interprete $\Gamma$ as the quantum effective action after the inclusion of all quantum effects, the resulting field equations $\partial \Gamma/\partial\tilde g^{\hat\mu\hat\nu}(x,y)=0,~\partial \Gamma/\partial \delta(x,y)=0$, are exact identities for the allowed states of the system.

\medskip\noindent
{\bf 1. Dependence of effective action on integration 

\hspace{0.2cm}constants}

Let us now evaluate $\Gamma$ for a finite dimensional subspace of metrics and scalar fields that we define in the following. Consider a given solution of the local field equations \eqref{18}-\eqref{21}, with singularities at $z=0$ and $z=\bar z$. We denote the corresponding functions by  $\rho_\gamma(z),\sigma_\gamma(z),\delta'_\gamma(z)$, with $\gamma=(E_0, L_0, \Lambda_0,\sigma_0)$ the remaining integration constants which specify the solution. (Two integration constants are fixed by the location of the singularities, and we recall that $\rho_0$ or, equivalently $\tilde C$ for $C\neq 0$, is fixed by eq. \eqref{21}.) Consider now a class of metrics given by the ansatz \eqref{14}, with $\sigma(z)=\frac{\hat\sigma}{\sigma_0}\sigma_\gamma(z),~\rho(z)=\frac{\hat\rho}{\rho_0}\rho_\gamma(z)$ and arbitrary $L$. This keeps the location of the singularities at $z=0$ and $z=\bar z$. We also keep the functions $\tilde g^{(4)}_{\mu\nu}(x)$ and $\bar g_{\bar\alpha\bar\beta}(y)$ fixed, where $\tilde g^{(4)}_{\mu\nu}$ corresponds to a given fixed value $\epsilon/\tilde L^2_0=\Lambda_0/L^2_0$ for given $\gamma$. Similarly, we consider scalar fields where $\delta'(z)$ is given by an arbitrary $E$ in eq. \eqref{18}. This subspace of field configurations is characterized by four variables $(E,L,\hat\sigma,\hat\rho)$ and we investigate the effective potential for these variables
\begin{equation}\label{P1}
W_\gamma(E,L.\hat\sigma,\hat\rho)=
\frac{M^{d-2}}{2}L^D
\int_z\sigma^2\rho^{\frac{D-1}{2}}
(\partial^{\hat \mu}\delta~\partial_{\hat \mu}\delta-\tilde R).
\end{equation}
The effective action depends on these variables via
\begin{equation}\label{P2}
\Gamma_\gamma(E,L,\hat\sigma,\hat\rho)=\int_x(\tilde g^{(4)})^{1/2}
W_\gamma(E,L,\hat\sigma,\hat\rho),
\end{equation}
and an extremum of the action corresponds to an extremum of $W_\gamma$. We employ a subscript $\gamma$ in order to recall that $W_\gamma$ depends implicitely on the chosen integration constants $E_0,L_0,\Lambda_0,\sigma_0$. For $E=E_0, L=L_0,\sigma=\sigma_0,~\hat\rho=\rho_0(E_0,L_0,\Lambda_0,\sigma_0)$, we know that $W$ must vanish identically due to eq. \eqref{13}.

Inserting the functions $\sigma_\gamma(z), \rho_\gamma(z),\delta'\gamma(z)$ one obtains
\begin{eqnarray}\label{P3}
W_\gamma&=&\frac12 M^{d-2}_d 
\left\{\frac{E^2\hat\sigma^2\hat\rho_\gamma^{\frac{D-1}{2}}L^{D-2}}
{E^2_0\sigma^2_0\rho_0^{\frac{D-1}{2}}}
\int_z\sigma^2_\gamma\rho_\gamma^{\frac{D-1}{2}}\delta'^2_\gamma\right.\nonumber\\
&&-\frac{4\Lambda_0}{L^2_0}
\frac{\hat\sigma\hat\rho^{\frac{D-1}{2}}L^D}{\sigma_0\rho_0^{\frac{D-1}{2}}}
\int_z\sigma_\gamma\rho_\gamma^{\frac{D-1}{2}}\nonumber\\
&&-\frac{\hat\sigma^2\hat\rho^{\frac{D-1}{2}}L^{D-2}}{\sigma^2_0\rho_0^{\frac{D-1}{2}}}
\int_z\sigma^2_\gamma\rho^{\frac{D-1}{2}}_\gamma\nonumber\\
&&\left.\left[\tilde R_{D,\gamma}+\left(\frac{\rho_0}{\hat\rho}-1\right)
\frac{(D-1)C}{\rho_\gamma}\right]\right\}.
\end{eqnarray}
Using eqs. \eqref{18} and \eqref{A7}
\begin{eqnarray}\label{P4}
\tilde R_{D,\gamma}&=&\delta'^2_\gamma-4\Lambda_0\sigma^{-1}_\gamma\nonumber\\
&=&E^2_0\sigma^{-4}_\gamma\rho^{-(D-1)}_\gamma-4\Lambda_0\sigma^{-1}_\gamma,
\end{eqnarray}
we find
\begin{eqnarray}\label{P5}
W_\gamma&=&\frac12 M^{d-2}_d
\frac{\hat\sigma^2\hat\rho^{\frac{D-1}{2}}L^{D-2}}
{\sigma^2_0\rho^{\frac{D-1}{2}}_0}\nonumber\\
&&\Big\{(E^2-E^2_0)\int_z\sigma^{-2}_\gamma\rho^{\frac{-(D-1)}{2}}_\gamma\nonumber\\
&&+4\Lambda_0\left(1-\frac{\sigma_0L^2}{\hat\sigma L^2_0}\right)
\int_z\sigma_\gamma\rho_\gamma^{\frac{D-1}{2}}\nonumber\\
&&+(D-1)C\left(1-\frac{\rho_0}{\hat\rho}\right)\int_z\sigma^2_\gamma\rho_\gamma^{\frac{D-3}{2}}\Big\}.
\end{eqnarray}
We recall that $\rho_0$ is fixed in terms of $E_0,L_0,\Lambda_0$ and $\sigma_0$ and observe that $W_\gamma$ indeed vanishes for $E=E_0,~\sigma_0/\hat\sigma=L^2_0/L^2,~\hat\rho=\rho_0$. 

\medskip\noindent
{\bf 2. Variation of internal volume implies $\Lambda = 0$}

One may expect that the higher dimensional solution $E=E_0,~L=L_0,~\hat\sigma=\sigma_0,~\hat\rho=\rho_0$ corresponds to an extremum of the action and therefore of $W_\gamma$. This is not the case. Let us first take $E=E_0,~\hat\sigma=\sigma_0,~\hat\rho=\rho_0$ and consider $W_\gamma$ as a function of the characteristic size of internal space $L$,
\begin{equation}\label{P6}
W_\gamma(L)=2\Lambda_0M^{d-2}_d\hat c_\gamma(L^{D-2}-L^DL^{-2}_0).
\end{equation}
The condition for an extremum at $L=L_0$
\begin{eqnarray}\label{P7}
\frac{\partial W_\gamma}{\partial L}_{|L_0}
&=&2\Lambda_0 M^{d-2}_d \hat c_\gamma
\big[(D-2)L^{D-3}-D L^{D-1}L^{-2}_0\big]\nonumber\\
&=&-4\Lambda_0M^{d-2}_d L^{D-3}_0\hat c_\gamma=0
\end{eqnarray}
is obeyed only for $\Lambda_0=0$. We infer that all solutions of the local higher dimensional field equations with $\Lambda_0\neq 0$ are not an extremum of the action. Extrema of the action \eqref{1} must have a vanishing effective four-dimensional cosmological constant $\Lambda_0=0$! It is precisely for this case that we have found a consistent dimensional reduction in the preceding section. 

The family of metrics with two singularities, where $\hat \sigma$ and $L$ are varied while $\hat\rho=\rho_0$ and $E=E_0$ are kept fixed, is perfectly acceptable. Since $\hat c_\gamma=\int_z\sigma_\gamma\rho_\gamma^{\frac{D-1}{2}}$ is finite, the potential $W_\gamma$ remains finite for this class of metrics. There is no doubt that the action should be an extremum within this family of metrics. For solutions with $\Lambda_0\neq 0$, there exist neighboring metrics for which the action per volume of four-dimensional space-time $W_\gamma$ changes by a finite amount, such that the solution is not an extremum. Therefore $\Lambda_0\neq 0$ is clearly not compatible with acceptable solutions. We conclude that for all extrema of the action with two singularities the effective four-dimensional cosmological constant vanishes without any tuning of parameters. 

\newpage\noindent
{\bf 3. Variation of integration constants for internal 

\hspace{0.2cm}geometry}

Let us next consider variations of $\hat\rho$, with $E=E_0,~\sigma=\sigma_0,~L=L_0$ fixed. Near the singularities one has
\begin{equation}\label{P7A}
\sigma^2\rho^{\frac{D-3}{2}}\sim z^{-2\eta+\frac{D-3}{2}\gamma}
\sim z^{\frac{D-3}{D-1}-\frac{4}{D-1}\eta}\sim z^{1-\gamma}.
\end{equation}
The $z$-integral of this expression remains finite near $z=0$, provided $\eta<\frac D2-1$ or $\gamma<2$. This holds for all $D\geq 3$. The class of metrics with varying $\hat\rho$ seems again to be acceptable. For $\Lambda_0=0,~E=E_0$, we remain with
\begin{eqnarray}\label{P8}
W_\gamma&=&\frac{D-1}{2}CM^{d-2}_d
\int_z\sigma^2_\gamma\rho^{\frac{D-3}{2}}_\gamma A,\nonumber\\
A&=&\frac{\hat\sigma^2}{\sigma^2_0}L^{D-2}
\left(\frac{\hat\rho^{\frac{D-1}{2}}}{\rho_0^{\frac{D-1}{2}}}
-\frac{\hat\rho^\frac{D-3}{2}}{\rho_0^{\frac{D-3}{2}}}\right).
\end{eqnarray}
For $C=0$ this does not yield an additional constant and we find that $W_\gamma$ vanishes identically. For $C\neq 0$, however, an extremum at $\hat\rho=\rho_0,~\hat\sigma=\sigma_0,~L=L_0$ requires $\sigma^2_0 L^{D-2}_0/\rho_0=0$. This is not consistent with our ansatz \eqref{14}. We conclude that for all extrema of the action with the ansatz \eqref{14} and two singularities, the space spanned by the coordinates $\bar y^\alpha$ must be Ricci-flat, i.e. $C=0,~\bar R_{\bar\alpha\bar\beta}=0$. We have seen, however, that no solutions with two singularities exist for $\Lambda=0, C=0$ - the exact singular solutions are the powerlike solutions \eqref{32}, \eqref{33}, \eqref{36} that do no lead to an acceptable four-dimensional gravity. Some ingredient is needed to stabilize $\hat\rho$ at some finite value, if the solutions with two singularities and $C\neq0$ are to play a role. One possibility are time-dependent cosmological solutions, where $\hat\rho(t)$ is finite for any finite time $t$. We will come back to this issue below. For the moment, we only note that the derivative 
\begin{equation}\label{P8A1}
\frac{\partial W_\gamma}{\partial\hat\rho}_{|\rho_0}
=\frac{D-1}{2}\tilde CM^4_d(M_dL)^{D-2}\int_z\sigma^2_\gamma\rho_\gamma^{\frac{D-3}{2}}
\end{equation}
gets small for small values of the integration constant $\tilde C$. 

We also should emphasize a difference between the physics associated to a variation of $\hat\rho$ and a variation of $L$. For fixed values $L=L_0,~\hat\rho=\rho_0,~E=E_0,~\hat\sigma=\sigma_0$, the dimensional reduction to an effective four-dimensional theory is valid whenever $\Lambda=0$, independently of the value of $C$. Contradictions arise only for $\Lambda\neq0$. In fact, we may evaluate $\Gamma$ for fixed $\sigma_\gamma,~\rho_\gamma,~\delta_\gamma$ only as a functional of an arbitrary metric $\tilde g^{(4)}_{\mu\nu}(x)$. Then $W_\gamma[\tilde g^{(4)}_{\mu\nu}]$ precisely corresponds to the effective four-dimensional Lagrangian and $\Gamma_\gamma[\tilde g^{(4)}_{\mu\nu}]$ to the effective four-dimensional action for the metric. The solution $\tilde g^{(4)}_{\mu\nu}(x)$ of the higher dimensional field equations is an extremum of $\Gamma_\gamma[\tilde g^{(4)}_{\mu\nu}]$ only for $\Lambda=0$. 

\medskip\noindent
{\bf 4. Variation of scalar integration constants}

Let us finally consider variations of $E$. In fact, the situation is more subtle if we keep the metric fixed, $L=L_0,~\hat\rho=\rho_0,~\hat\sigma=\sigma_0$, but vary the scalar field with $E\neq E_0$.  For a finite coefficient of the term $\sim E^2$ in $W_\gamma$ one concludes immediately that the only extremum with $\partial W_\gamma/\partial E=0$ occurs for $E=0$, and therefore infer $E_0=0$. However, the coefficient in the first term in eq. \eqref{P5} is not finite if singularities are present, since
\begin{equation}\label{P8A}
\sigma^{-2}\rho^{-\frac{D-1}{2}}\sim z^{2\eta-\frac{D-1}{2}\gamma}\sim z^{-1}. 
\end{equation}
We conclude that $E\neq E_0$ does not describe an acceptable field configuration with finite $W_\gamma$. For a given metric only the choice $E=E_0$ ensures that $W_\gamma$ remains finite. 

This issue may be better understood if we write the piece in the action containing the scalar field as
\begin{eqnarray}\label{P9}
\Gamma_\delta&=&\frac{M^{d-2}_d}{2}\int \tilde g^{1/2}\partial^{\hat\mu}\delta\partial_{\hat\mu}\delta\\
&=&-\frac{M^{d-2}_d}{2}
\int\big\{\tilde g^{1/2}\delta\hat D^2\delta-\partial_{\hat\mu}(\tilde g^{1/2}\delta\partial^{\hat\mu}\delta)\big\}.\nonumber
\end{eqnarray}
The first term vanishes for any solution of the field equations. However, there is a second ``boundary term'', which reads for our local solution
\begin{eqnarray}\label{P10}
\Gamma_\delta&=&\frac{M^{d-2}_d}{2}
L^{D-2}\int_x(\tilde g^{(4)})^{1/2}
\int_z\partial_z(\sigma^2\rho^{\frac{D-1}{2}}\delta\delta')\nonumber\\
&=&\frac{M^{d-2}_d}{2}L^{D-2}
E\int_x(\tilde g^{(4)})^{1/2}
\big(\delta(\bar z)-\delta(0)\big).
\end{eqnarray}
More precisely, $\delta(0)$ stands for $\delta(z\to 0_+)$ and $\delta(\bar z)$ for $\delta(z\to\bar z_-)$. In fact, near the singularities one has $\delta'\sim z^{-1}$ and $\delta'\sim(\bar z-z)^{-1}$, respectively. Therefore $\delta$ diverges logarithmically to $-\infty$ for $z\to 0_+$, and to $+\infty$ for $z\to\bar z_-$, if $E>0$ (and inversely for $E<0$). The difference $\delta(\bar z)-\delta(0)$ therefore diverges and $\Gamma_\delta$ is not well defined. On the other hand, the curvature part of the action, $\Gamma_R$, contains a similar divergence. Indeed, for any solution of the field equations one has
\begin{equation}\label{124A}
\tilde R=(\delta')^2
\end{equation}
and therefore
\begin{equation}\label{124B}
\sigma^2\rho^{\frac{D-1}{2}}\tilde R=E_0\delta'.
\end{equation}
Insertion into $\Gamma_R$ yields
\begin{eqnarray}\label{124C}
\Gamma_R&=&-\frac{M^2_d}{2}\int\tilde g^{1/2}\tilde R\\
&=&-\frac{M^{d-2}}{2}V_{D-1}L^{D-2}E_0\int_x(\tilde g^{(4)})^{1/2}\big(\delta(\bar z)-\delta(0)\big)\nonumber
\end{eqnarray}
such that for $E=E_0$ one finds $\Gamma_R=-\Gamma_\delta$. 

If we restrict the allowed field configurations to the ones where $\Gamma/V_4$ (with $V_4$ the volume of four-dimensional space-time) remains finite, the divergences in $\Gamma_R$ and $\Gamma_\delta$ must cancel. This fixes the allowed value of $E$ as a function of the singularity in the metric. For solutions of the field equations this requires $E=E_0$ for given $\sigma_\gamma,\rho_\gamma$. At this stage we are then not allowed to vary $E$ and should consider $W_\gamma$ only as a function of $L,~\hat\sigma,~\hat\rho$, omitting the first term in eq. \eqref{P5}. 

Instead of varying $\delta$ by a multiplicative constant at fixed $\tilde g_{\hat\mu\hat\nu}$, we may alternatively consider a shift of $\xi$ by an additive constant, $\xi\to\xi+\epsilon$, at fixed $\hat g_{\hat\mu\hat\nu}$. We first verify that the extrema of the action \eqref{10a} coincide with the extrema of the original action \eqref{1}. This is not automatic since the transformation has neglected a boundary term
\begin{eqnarray}\label{P11}
\Delta \Gamma&=&\frac{g_d}{2}\int\tilde g^{1/2}\xi^2 w^{d-2}\hat D^2\ln w\\
&=&-\frac{g_d}{d-2}M^{d-2}_d\int\partial_{\hat\mu}(\tilde g^{1/2}\partial^{\hat\mu}\ln\xi)\sim
\int_z\partial_z(\sigma^2\rho^{\frac{D-1}{2}}\delta').\nonumber
\end{eqnarray}
However, this term vanishes identically for the solutions of the field equations such that the extrema are the same for eqs. \eqref{1} and \eqref{10a}. For a constant infinitesimal shift $\epsilon$ the variation of the action \eqref{1} reads simply
\begin{equation}\label{124D}
\delta \Gamma=-\int \hat g^{1/2}\xi\hat R\epsilon=
-\int_x(\tilde g^{(4)})^{1/2}\int_z w^d\sigma^2
\rho^{\frac{D-1}{2}}\xi\hat R\epsilon.
\end{equation}

If a constant shift $\epsilon$ would correspond to a finite variation $\delta \Gamma/V_4$, i.e. for finite $\int_z w^d\sigma^2\rho^{\frac{D-1}{2}}\xi\hat R$, the extremum condition $\delta \Gamma=0$ would imply $\hat R=0$. (Note $\xi=\exp(c_\delta\delta)\geq 0$ for positive $c_\delta=[\zeta+4f_d/(d-2)^2]^{1/2}$ and $w,\sigma,\rho>0$.) Combining eqs. \eqref{2}, \eqref{3A} one finds for all solutions of the field equations
\begin{equation}\label{124E}
\hat R=\zeta\partial_{\hat \mu} \ln\xi\partial_{\hat\nu}\ln\xi\hat g^{\hat\mu\hat\nu}
=\zeta w^{-2}c^2_\delta(\delta')^2,
\end{equation}
such that $\hat R=0$ corresponds to $E_0=0$. However, with
\begin{equation}\label{124F}
y=w^d\sigma^2\rho^{\frac{D-1}{2}}\xi\hat R=\zeta M^{d-2}_d
c^2_\delta E^2_0\xi^{-1}\sigma^{-2}\rho^{-\frac{D-1}{2}}
\end{equation}
one finds for $E_0>0$, with $\alpha_y>0,\beta_y>0$
\begin{equation}\label{124G}
y(z\to 0)\sim z^{-1-\alpha_y},y(z\to \bar z)\sim (\bar z-z)^{-1+\beta_y}.
\end{equation}
Therefore $\int_zy$ converges for $z\to\bar z$, but diverges for $z\to 0$. Even though the variation at constant $\hat g_{\hat\mu\hat\nu}$ shows a different behavior at the singularities as compared to the one with fixed $\tilde g_{\hat\mu\hat\nu}$, it still changes $\Gamma/V_4$ by an infinite amount.

Nevertheless, there are other variations of $\delta$ which leave $\Gamma/V_4$ finite, but imply $E_0=0$ for the extremum condition. Consider a $z$-dependent shift $\delta\to\delta +\epsilon(z)$, with fixed $\tilde g_{\hat\mu\hat\nu}$. The variation of the action \eqref{10a} becomes
\begin{eqnarray}\label{124H}
\delta \Gamma&=&M^{d-2}_d\int\tilde g^{1/2}\partial^{\hat\mu}\delta\partial_{\hat\mu}\epsilon\\
&=&M^{d-2}_d\big\{\int\partial_{\hat \mu} (\epsilon\tilde g^{1/2}\partial^{\hat \mu}\delta)
-\int\epsilon\tilde g^{1/2}\hat D^2\delta\big\}.\nonumber
\end{eqnarray}
The second contribution vanishes by virtue of the field equation \eqref{11aneu} such that 
\begin{eqnarray}\label{124I}
\delta \Gamma&=&M^{d-2}_d\int_x(\tilde g^{(4)})^{1/2}
\int_z\partial_z(\epsilon\sigma^2\rho^{\frac{D-1}{2}}\delta')\nonumber\\
&=&M^{d-2}_dE_0\int_x(\tilde g^{(4)})^{1/2}\int_z\partial_z\epsilon(z).
\end{eqnarray}
If we take $\epsilon(z)=\kappa z$ we obtain for the spaces with two singularities at $z=0$ and $z=\bar z$ a finite value $\int_z\epsilon=V_{D-1}\kappa \bar z$. The extremum condition with respect to such a variation, $\delta \Gamma=0$, implies $E_0=0$. If we admit solutions of the field equations where $\delta$ diverges at the singularities, there seems to be no reason to exclude the configurations where $\delta$ is shifted by a finite amount, as $\delta\to\delta+\kappa z$. More generally, it is always possible to find $\epsilon(z)$ with $\int_z\partial_z\epsilon\neq 0$. We infer that all extrema of the action have $E=0$ and therefore $z$-independent values of the scalar field, $\delta=\delta_0,\xi=\xi_0$. In this case $\delta \Gamma=0$ is obeyed for arbitrary $\epsilon$, cf. eq. \eqref{124H}.

\medskip\noindent
{\bf 5. Extrema of effective action}

We conclude that the extrema of the action require for quasistatic solutions with a warped geometry \eqref{14} the choice of integration constants $E=0,\Lambda=0,C=0$. The remaining integration constants $\sigma_0,\rho_0,L$ correspond to trivial rescalings of the coordinates in this case. The singular solutions with warping are given by the powerlike solution \eqref{32}, with $\eta$ and $\gamma$ determined by eqs. \eqref{36A}, \eqref{33}. They do not lead to a finite four-dimensional Planck mass. If we restrict the range of $z$ to $0<z<z_{max}$ the effective four-dimensional Planck mass diverges $\chi^2\sim z^2_{max}$. We may imagine a cosmology where this particular static solution is approached asymptotically, with an effective $z_{max}$ increasing with time. However, for a realistic particle physics the typical mass scale for the Kaluza-Klein modes has also to increase with $z_{max}$, and the dimensionless couplings should remain essentially static despite the increase of $z_{max}$. It may be difficult to achieve these requirements. 

The absence of solutions with $\Lambda =C=E=0$ and finite $\chi$ is a particularity of our limitation to an $SO(D)$ isometry. Other solutions with a different isometry group are compatible with $\Lambda =C=E=0$ and lead nevertheless to a consistent four-dimensional gravity with finite $\chi$.
For the ansatz \eqref{14}  the extrema with a direct product structure $\sigma(z)=\sigma_0,\rho(z)=\rho_0$, with $\bar g_{\bar\alpha\bar\beta}(y)$ the metric of a Ricci-flat $D-1$-dimensional space with finite volume $V_{D-1}$, are consistent. This also holds if $\hat g_{\alpha\beta}(y,z)$ describes a Ricci-flat $D$-dimensional geometry. Such spaces are indeed extrema of $\Gamma$. We conclude that internal geometries corresponding to extrema of $W$ in eqs. \eqref{2D}, \eqref{2E} typically exist. They single out a vanishing cosmological constant $\Lambda = 0$.

This result has far reaching consequences. In particular, it explains within our model why asymptotic cosmological solutions lead to a vanishing cosmological constant. We should therefore understand better why the local solutions of the field equations are not automatically extrema of the action. The basic reason is that the local field equations \eqref{11aneu}, \eqref{12} are obtained from  {\em local} variations of the action. Local variations $\delta\tilde g^{\hat\mu\hat\nu}(x,y)$ typically vanish outside some region in higher dimensional space-time. Since any extrema of $\Gamma$ must also be extrema with respect to local variations, it is a necessary condition for any state that the field equations are obeyed. For compact internal spaces with a regular warp factor, this condition is also sufficient. In presence of singularities it is not. 

In other words, in addition to the local variations we have also to consider variations of the $d$-dimensional fields that are not local in $d$-dimensional space-time. For example, a variation of $L$ changes the internal metric for all values of $z$, not only in a local region. This holds even if $L(x)$ is local in four-dimensional space-time. The extremum condition for $\Gamma$ under a variation of $L$ is not contained in the field equations. It has to be imposed in addition.

We therefore can device the following strategy for finding the extrema of $\Gamma$. First one solves the field equations. The most general solution consistent with a given symmetry has typically a number of integration constants $\gamma_j$. In the second step, one may evaluate $\Gamma$ as a function of the integration constants by inserting the corresponding solutions of the field equations for some fixed ``trial values'' $\gamma=\gamma_{j,0}$. This leads to a potential $W_\gamma(\gamma_j)$ for the integration constants. The third step determines the extrema of $W$. These are then the extrema of the action. The resulting extremum condition typically fixes part of the integration constants $\gamma_j$. Alternatively, one may perform dimensional reduction to an effective four-dimensional theory and use the requirement that an extremum of $\Gamma$ has to obey the effective four-dimensional field equations. 

For compact internal spaces with a regular warp factor the situation is special. Now part of the integration constants $\gamma_j$ are fixed by the conditions of regularity. The solutions of the field equations are automatically extrema of $W_\gamma$ in this case. This follows from the absence of boundary terms, such that the local variations are sufficient to find the extrema of $\Gamma$. We conclude that compact regular solutions are always acceptable states. There are, however, additional possibilities with singularities. In this case the conditions on the integration constants $\gamma_j$ for the regularity are replaced by the extremum conditions for $W_\gamma$. 

\section{General dilatation symmetric effective action}
\label{generaldilatation}

In this section we generalize the discussion of the possible extrema of $\Gamma$ in sect. \ref{potentialfor}. We add to the effective action \eqref{1} the most general dilatation symmetric pure gravitational part,

\be\label{138A}
\Gamma= \int \hat g^{1/2} \left\{-\frac12\xi^2\hat R+\frac \zeta2\partial^{\hat\mu}\xi\partial_{\hat\mu}\xi
+F(\hat R\hmnrs)\right\}.
\ee
We only assume here that $F$ is dilatation invariant and depends only on the metric and its derivatives, being a scalar under general coordinate transformations. We do not need to restrict the discussion to a polynomial form of $F$. 

\medskip\noindent
{\bf 1. General solutions}

We also consider much more general solutions as in the preceding sections by investigating all possible quasi-static solutions with a warped metric (4) and $\xi = \xi(y)$. The detailed form of the solutions will not be important for the general arguments of this section. We only need the observation that there are families of higher dimensional solutions with free integrations constants as we have encountered in the preceding sections. The overall scale of $\xi$ and the characteristic length scale $l$ of internal space count among these integration constants. We assume that the solutions are of the type admitting dimensional reduction to an effective four-dimensional theory of gravity. Within this very general setting we will find a class of solutions in the ''flat phase'' with vanishing four-dimensional cosmological constant $\Lambda = 0$. These solutions are extrema of $\Gamma$ and are stable. In contrast no stable extrema of $\Gamma$ with $\Lambda \neq 0$ exist if $\xi \neq  0$. These statements hold independently of the values of the couplings that parameterize $F$.

We will work directly with the action\eqref{138A} without performing a Weyl scaling. Nevertheless, direct contact with the results of the preceding sections can be easily made. In this case we use the ansatz \eqref{14} and employ the
simple relation \eqref{4}, \eqref{7} between $\hat g_{\hat\mu\hat\nu}$ and $\tilde g_{\hat\mu\hat\nu}$ which amounts to appropriate multiplicative rescalings for the metric components. We can use 
\begin{equation}\label{XS1}
g^{(4)}_{\mu\nu}=M^2_d\xi^{-\frac{4}{d-2}}\tilde g^{(4)}_{\mu\nu}
\end{equation}
and absorb the multiplicative factor between $\tilde g_{\alpha\beta}$ and $\hat g_{\alpha\beta}$ in $L^2$, which is not determined by the higher dimensional solution anyhow. 

\medskip\noindent
{\bf 2. Radion and dilaton}

We will perform dimensional reduction to an effective four-dimensional theory, expanding the effective action \eqref{138A} around a solution of the higher dimensional field equations. The effective four-dimensional action $\Gamma^{(4)}$ will depend on the values of the integration constants which characterize the particular solution around which we expand. Among these integration constants we only keep $\xi$ and $l$ as parameters of the four-dimensional theory. They play the role of four-dimensional scalar fields and can be associated with the dilaton and the radion. The quasi-static solutions must be extrema of $\Gamma^{(4)}$ with respect to variations of $\xi$ and $l$. For the other integration constants we assume that they have fixed values which correspond to a (partial) extremum of $\Gamma$ and $\Gamma^{(4)}$. (This means that the derivatives of $W_\gamma$ in sect.\ref{potentialfor} with respect to these parameters vanish.) This procedure is illustrated by the detailed discussion in the preceding sections. In this case we insert for $\bar g_{\bar\alpha\bar\beta}(\bar y)~,~\rho(z)~,~\sigma(z)~,~\xi(z)$ one of the solutions of the higher dimensional field equations, as given by a set of integration constants $\gamma=(\Lambda,\dots)$. These solutions do not fix $L$ and the normalization of $\xi$, such that we can indeed retain two scalar degrees of freedom.

The strong result of this section that only solutions with $\Lambda=0$ are possible stable extrema of $\Gamma$ will be based on the extremum conditions for $\xi$ and $l$ as well as an investigation of stability of possible extrema. For the discussion of stability we will need the second variations of $\Gamma^{(4)}$ with respect to $\xi$ and $l$.

The general form of the four-dimensional action can be written as
\begin{equation}\label{8Ba}
\Gamma^{(4)}=\int_x(g^{(4)})^{1/2}\left\{V-\frac12 \chi^2R^{(4)}+\dots\right\}.
\end{equation}
We regard $\Gamma^{(4)}$ as an expansion in powers of the four-dimensional curvature tensor and the dots denote terms involving higher powers of it. The potential $V$ plays the role of the four-dimensional effective potential for $\xi$ and $l$, where we recall that all quantum fluctuations are already included. It is given by

\begin{equation}\label{XS2}
V=\int_y(g^{(D)})^{1/2}\sigma^2\Big\{-\frac12\xi^2R^{(\textup{int})}+
\frac\zeta2\partial^\alpha\xi\partial_\alpha\xi+
F(R^{(\textup{int})}_{\hat\mu\hat\nu\hat\rho\hat\sigma})\Big\}.
\end{equation}
Here $R^{(\textup{int})}$ and $F(R^{(\textup{int})}_{\hat\mu\hat\nu\hat\rho\hat\sigma})$ are evaluated by inserting for the $d$-dimensional curvature tensor the functions $g^{(D)}_{\alpha\beta}(y),\sigma(y),\xi(y)$, which correspond to a particular solution of the $d$-dimensional field equation, while the four-dimensional metric is kept flat, $g^{(4)}_{\mu\nu}=\eta_{\mu\nu}$. Quantities like $R^{(\textup{int})}$ and therefore $V$ will depend on the choice of the integration constants which characterize the higher dimensional solution. 

On the other hand, one finds for the effective gravitational constant
\begin{equation}\label{8D}
\chi^2=\int_y(g^{(D)})^{1/2}\sigma\{\xi^2-2G\},
\end{equation}
where we use
\begin{equation}\label{XS3}
\hat R=R^{(\textup{int})}+R^{(4)}/\sigma.
\end{equation}
Here $G$ obtains from the first order term in an expansion of $F$ in $R^{(4)}/\sigma$,
\begin{equation}\label{8E}
F=F(R^{(int)}_{\hat\mu\hat\nu\hat\rho\hat\sigma})+GR^{(4)}/\sigma+\dots
\end{equation}
For the example $F=\tau\hat R^{\frac{d}{2}}$ one has
\begin{equation}\label{8F}
F=\tau(R^{(int)})^{\frac d2}+\frac{\tau d}{2}(R^{(int)})^{\frac d2-1}R^{(4)}/\sigma+\dots
\end{equation}
such that
\begin{equation}\label{190A}
G=\frac{\tau d}{2}(R^{(\textup{int})})^{\frac d2-1}.
\end{equation}

We can now define the characteristic length scale $l$ for the internal geometry by
\begin{equation}\label{YA}
\int_y(g^{(D)})^{1/2}\sigma^2=l^D.
\end{equation}
Similarly, we define the characteristic scale $\bar\xi$ for the scalar field by
\begin{equation}\label{YB}
\int_y(g^{(D)})^{1/2}\sigma\xi^2=l^D\bar\xi^2.
\end{equation}
This definition is obvious for a constant $\xi$ and $\sigma=1$, but can now be applied for arbitrary configurations, including $y$-dependent scalar configurations with $\partial_\alpha\xi\neq 0$ and warping.

\medskip\noindent
{\bf 3. Effective potential and Planck mass}

We define dimensionless constants 
\begin{eqnarray}\label{Y6}
\tilde F&=&l^4\int_y(g^{(D)})^{1/2}\sigma^2F(R^{(\textup{int})}_{\hat\mu\hat\nu\hat\rho\hat\sigma}),\nonumber\\
\tilde G&=&l^2\int_y(g^{(D)})^{1/2}\sigma G,
\end{eqnarray}
and observe that $\tilde F$ and $\tilde G$ do not change under a rescaling of the internal metric. They can therefore be evaluated for a fixed $l_0$. (For the definition of $\tilde F$ and $\tilde G$ it is understood that the same metric is used for all quantities on the r.h.s..) Similarly, we define
\begin{equation}\label{XS4}
\tilde Q=\frac12\bar\xi^{-2}l^{2-D}\int_y(g^{(D)})^{1/2}\sigma^2(\zeta\partial^\alpha\xi
\partial_\alpha\xi-\xi^2 R^{(\textup{int})}).
\end{equation}
(Again, the definition \eqref{XS4} is understood in the sense that the solution for $\xi(y)$ is also used for the computation of $\bar\xi$ on the r.h.s., and similar  for the scaling of the internal metric.) In terms of these constants we can write

\be\label{150A}
V=\tilde Q \bar\xi^{2} l^{D-2} +\tilde F l^{-4},
\ee
and
\be\label{150B}
\chi^2 =l^D \bar\xi^{2} - 2 \tilde G l^{-2}.
\ee
The constants $\tilde Q,\tilde F, \tilde G$ have to be evaluated for a particular solution of the higher dimensional field equations. In general, they will depend on the integration constants $\gamma$ used for this solution. 

We further restrict the solutions to obey the condition
\begin{equation}\label{XS5}
\int_y\partial_\alpha(\hat g^{1/2}\xi\partial^\alpha\xi)=0.
\end{equation}
This condition has to be met for any extremum of $\Gamma$. Indeed, we may consider an infinitesimal variation $\xi(y,x)\to\xi(y,x)+\epsilon(y,x)$. This results in a variation of $\Gamma$ \eqref{138A}
\begin{eqnarray}\label{XS6}
\delta\Gamma&=&\int_{\hat x}\hat g^{1/2}
\{\zeta\partial_{\hat\mu}\epsilon\partial^{\hat\mu}\xi-\epsilon\xi\hat R\}\\
&=&\zeta\int_{\hat x}\partial_{\hat\mu}(\hat g^{1/2}\epsilon\partial^{\hat\mu}\xi)-
\int_{\hat x}\hat g^{1/2}\epsilon(\zeta\hat D^2\xi+\hat R\xi).\nonumber
\end{eqnarray}
The second term vanishes by virtue of the field equation \eqref{2a}, such that an extremum of $\Gamma~(\delta\Gamma=0)$ requires
\begin{equation}\label{XS7}
\int_{\hat x}\partial_{\hat\mu}(\hat g^{1/2}\epsilon\partial^{\hat\mu}\xi)=0
\end{equation}
for all acceptable variations $\epsilon$. For $\xi$ depending only on $y$ and the choice $\epsilon=\bar\epsilon(x)\xi(y)$, with $\bar\epsilon(x)$ non-vanishing inside some local region in four-dimensional spacetime, an extremum implies the condition \eqref{XS5}. 

We can use the extremum condition \eqref{XS7} in order to bring $\tilde Q$ into an intuitive form. Any $d$-dimensional solution must obey the scalar field equation \eqref{2} which does not depend on the term $F$. For quasistatic solutions $(\partial_\mu\xi=0)$ we can combine the field equation \eqref{2} and the extremum condition \eqref{XS5} in order to bring $\tilde Q$ to the form
\begin{equation}\label{XS8}
\tilde Q=\frac12\bar\xi^{-2}l^{2-D}\int_y(g^D)^{1/2}\sigma^2\xi^2
(\hat R-R^{(\textup{int})}).
\end{equation}
Both $\hat R$ and $R^{(\textup{int})}$ have to be evaluated for a solution of the higher dimensional field equations. (For the warped solutions of the preceding sections they are characterized by the integration constants $(\Lambda_0,\sigma_0,L_0)$, while  $E_0=0$ is required by eq. \eqref{XS5}.) 

\medskip\noindent
{\bf 4. Solutions with $\Lambda=0$}

Our strategy is to investigate the possible values of $\tilde F$, $\tilde G$ and $\tilde Q$ for which a stable quasistatic solution of the field equations derived from $\Gamma^{(4)}$  \eqref{8B} exists. We will find that for $\bar\xi \neq 0$ the only possibility is $\tilde Q = \tilde F = 0$. Consistent stable quasi-static solutions therefore exist whenever the integration constants of the $d$-dimensional solution can be chosen such that $\tilde Q = \tilde F = 0$, while the derivative of $W_\gamma$ with respect to all integration constants except $\bar\xi$ and $l$ vanishes. We will see that these solutions all have $\Lambda = 0$ and indeed correspond to an extremum of $\Gamma$. We recall that the four-dimensional cosmological constant is one of the integration constants. The $d$-dimensional solution around which dimensional reduction is performed will therefore depend on some ``trial value'' $\Lambda_0$. Of course, consistency requires that the value of $\Lambda$ found from the quasistatic extremum of eq. \eqref {8B} coincides with $\Lambda_0$. We will investigate separately the cases $\Lambda_0 = 0$ and $\Lambda_0 \neq 0$. They correspond to the two different phases for the possible solutions.

Let us first expand around a solution with $\Lambda_0=0$. In this case one has $R^{(\textup{int})}(\Lambda_0=0)=\hat R(\Lambda_0=0)$ and therefore concludes $\tilde Q=0$. For this choice of integration constants the potential \eqref{XS2} is given by 
\begin{equation}\label{xx}
V=\tilde Fl^{-4}.
\end{equation}
There are two alternative situations, according to the existence of a solution of the higher dimensional field equations with $\tilde F=0$ or not. Consider first the case where a suitable choice of integration constants allows for a solution with $\tilde F=0$, while the condition \eqref{XS5} is also obeyed. This seems to be a generic situation, since there are typically several integration constants $\gamma_i$ on which $\tilde F(\gamma_i)$ depends, and only two are fixed by $\Lambda_0=0$ and the condition \eqref{XS5}. (For the warped solutions in the preceding section two free integration constants $L_0,\sigma_0$ remain after $\Lambda_0=0,E_0=0$ are fixed.) If $\tilde F(\gamma_i)=0$ is possible we can perform dimensional reduction to an effective four-dimensional model by expanding around the higher dimensional solution with integration constants $\gamma^{(0)}_i$ chosen such that $\tilde F(\gamma^{(0)}_i)=0,\tilde Q(\gamma^{(0)}_i)=0$. The effective potential is then found to be independent of $l$ and $\bar \xi$,
\begin{equation}\label{MA}
V(l,\bar \xi)=0.
\end{equation}
The field equations for $l$ and $\bar \xi$ are obeyed, and the solution of the gravitational field equations, $g^{(4)}_{\mu\nu}=\eta_{\mu\nu}$, is consistent with a vanishing cosmological constant, $\Lambda_0=0$. Such a quasistatic configuration therefore solves both the higher dimensional and the four-dimensional field equations. For this class of solutions the ``tuning'' of the cosmological constant to zero happens independently of the values of the higher dimensional couplings, i.e. $\zeta$ and the dimensionless couplings parameterizing $F$. 

The solutions with $\Lambda = 0$ define the flat phase of the possible extrema of $\Gamma$. They obey $W_0=0$ for the discussion in the introduction, eqs. \eqref{2D} - \eqref{8B}. For $W_0=0$ we can turn eq. \eqref{8A} around and infer from $\delta\Gamma=0$ that $\delta W$ must vanish, such that the solution corresponds to an extremum of $W$. Extrema in the flat phase always exist for the general dilatation symmetric effective action \eqref{138A}. Indeed, we may consider the direct product of four-dimensional Minkowski space and a $D$-dimensional torus ${\cal M}^4 \times {\cal T}^D$, accompanied by a constant value for $\xi$. For a finite volume of the torus this leads to consistent four-dimensional gravity with finite nonzero ${\chi}$. One finds $\tilde F=\tilde G=\tilde Q=0$. Since ${\cal M}^4\times {\cal T}^D$ is a regular space the solution of the $d$-dimensional field equations is automatically an extremum of $\Gamma$. The interesting question concerns the ``size'' or ``extension'' of the flat phase, i.e. a classification which non-trivial solutions belong to it. If for a given ansatz one finds no integration constants which extremize  $W_\gamma$ and allow for $\tilde F=0$, or if such solutions do not lead to a finite nonzero $\chi^2$, this only means that no solutions consistent with the ansatz belong to the flat phase. This is what we have encountered in the preceding sections for the most general ansatz with $SO(D)$-symmetry, where finite ${\chi^2}$ has not been realized.

\medskip\noindent
{\bf 5. Possible solutions with $\Lambda\neq 0$}

Let us next investigate possible additional extrema belonging to the non-flat phase.  They would have a non-vanishing cosmological constant $\Lambda\neq 0$. We will find that stable solutions are not possible for $\bar\xi\neq 0$. This finding is crucial for an understanding why ${\Lambda=0}$ is singled out. Let us suppose that within a given ansatz one finds a choice of integration constants such that a higher dimensional solution with $\Lambda_0=0$ exists. One may ask what happens if we perform a dimensional reduction by expanding around some neighboring higher dimensional solution, with integration constants such that $\Lambda_0\neq 0$. We will show that this leads to instabilities, indicating that such higher dimensional solutions cannot be stable extrema of the action. 

For a general higher dimensional solution with integration constant $\Lambda_0$, given by $R^{(4)}=4\Lambda_0$, eqs. \eqref{XS8} and \eqref{XS3} imply 
\begin{equation}\label{T1}
V=\int_y(g^{(D)})^{1/2}
\left\{2\Lambda_0\xi^2\sigma+\sigma^2 F(R^{(int)}_{\hat\mu\hat\nu\hat\rho\hat\sigma})\right\}.
\end{equation}
We can directly express $\tilde Q$ in terms of the integration constant $\Lambda_0$ and the scale $l_0$ using eqs. \eqref{XS8}, \eqref{YB}
\begin{equation}\label{T2}
\tilde Q=2\Lambda_0l^2_0.
\end{equation}
Here $l_0$ is the ``trial scale'' for which $\tilde Q$  is evaluated. (If we choose a different $l_0$ also $\Lambda_0$ should be changed accordingly, since for any higher dimensional solutions $\hat R, R^{int}\sim l^{-2}_0$. Therefore $\tilde Q$ should be considered as a constant and $l_0$ should be kept fixed if we later vary the radion field $l$ away from the space of $d$-dimensional solutions. The independent integration constant is the dimensionless combination $\Lambda_0l^2_0$)

\medskip\noindent
{\bf 6. Maximally symmetric four-dimensional 

\hspace{0.2cm}geometry}

For the solutions of the gravitational equations derived from the effective action \eqref{8B} we concentrate on spaces with maximal symmetry 
\be\label{G18a}
R^{(4)}=4\Lambda.
\ee
Inserting this solution into $\Gamma^{(4)}$ \eqref{8B} we define the ``cosmological potential'' 
\be\label{G18b}
W(\bar\xi,l)=V-2\Lambda\chi^2.
\ee
It differs from the effective potential $V$ by a term proportional to the cosmological constant. Extrema of $\Gamma^{(4)}$ correspond to extrema of $W$ with respect to variations of $\bar\xi$ and $l$, not to extrema of $V$. (Only for $\Lambda=0$ the two potentials $W$ and $V$ coincide.) Inserting eqs. \eqref{150A}, \eqref{150B}, \eqref{T2} we obtain the explicit form of the cosmological potential 
\begin{equation}\label{T3}
W=2\Lambda_0l^2_0\bar\xi^2l^{D-2}+\tilde F l^{-4}-2\Lambda
(\bar\xi^2l^D-2\tilde Gl^{-2}).
\end{equation}

We concentrate on solutions with $\bar\xi\neq 0$. The field equation for $\bar\xi$ has the the simple solution
\begin{equation}\label{214A}
\Lambda l^2=\Lambda_0l^2_0.
\end{equation}
In this respect the four-dimensional solution is consistent with the ansatz used for the dimensional reduction. The second condition for an extremum, namely $\partial W/\partial l=0$, reads
\begin{equation}\label{UA}
(D-2)\Lambda_0l^2_0\bar\xi^{-2}l^{D-2}-2\tilde F l^{-4}-D\Lambda\bar\xi^{2}l^D-4\tilde G\Lambda l^{-2}=0.
\end{equation}
Insertion of $\Lambda l^2=\Lambda_0l^2_0$ implies the extremum condition
\begin{equation}\label{UB}
\Lambda_0\bar\xi^2l^D_0=-(\tilde F+2\tilde G\Lambda_0l^2_0)l^{-4}_0.
\end{equation}
Depending on the values of $\tilde F$ and $\tilde G$ this equation can have solutions with $\Lambda_0\neq 0$, such that corresponding extrema of $\Gamma$ may exist.

\medskip\noindent
{\bf 7. Stability of solutions}

For $\Lambda_0<0$ the leading term in $W$ \eqref{T3} for $\bar \xi^2\to\infty,l\to 0$ goes to minus infinity. A stability analysis reveals that possible extrema with $\Lambda_0<0$ are unstable. An instability also occurs for extrema with $\Lambda_0>0$ which become allowed for $\tilde G<0$. The matrix of second derivatives, 
\be\label{162A}
\hat m^2_{ij}=\frac12
\frac{\partial^2 W}{\partial x_i\partial x_j}~,~(x_1,x_2)=(l,\bar\xi), 
\ee
evaluated for the extremum conditions \eqref{214A}, \eqref{UB}, reads
\begin{equation}\label{UC}
\hat m^2=4\left(\begin{array}{ccc}
\tilde F l^{-6}_0-D\Lambda_0\bar \xi^2l^{D-2}_0&,&-\Lambda_0\bar\xi l^{D-1}_0\\
-\Lambda_0\bar\xi l^{D-1}_0&,&0
\end{array}\right).
\end{equation}
Due to the vanishing of $\hat m^2_{22}$ stability is only possible if the off diagonal elements vanish. For nonzero $\bar\xi$ and finite $l_0$ this is possible only for $\Lambda_0=0$. In this case the extremum condition \eqref{UB} requires $\tilde F=0$, and $\hat m^2$ vanishes identically at the extremum. 

In presence of the pure curvature term $\sim F$ in the effective action we conclude that extrema with $\Lambda_0\neq 0$ cannot be excluded. (In contrast to the extrema with $\Lambda_0=0$ we have not yet found an explicit example, however.) This contrasts with the situation for $F=0$ where the extremum  conditions lead to contradictions between the higher-dimensional and the four-dimensional field equations unless $\Lambda_0=0$. However, all possible extrema with $\bar\xi>0$, finite $l$ and $\Lambda_0\neq 0$ turn out to be unstable. Again, the solutions with $\Lambda_0=0$ are singled out.

The stability of an extremum of $\Gamma$ is crucial for its relevance for the asymptotic cosmological solution as time goes to infinity. In our picture a runaway cosmological solution approaches the field region dominated by the ultraviolet fixed point only asymptotically. Any instability of the solution will deviate a given solution from a trajectory towards the fixed point. We conclude that in presence of possible stable and unstable fixed point solutions only the stable ones have a chance to reached. In our case this type of cosmology can asymptotically only approach the solutions with $\Lambda=0$.

\section{GRAVITY WITHOUT DILATON}
\label{gravitywithout}

In this section we discuss the case of a dilatation symmetric effective action without a dilaton field. The effective action \eqref{138A} contains then only the purely gravitational part $\sim  F$. The relevant solutions will simultaneously cover the solutions with $\xi =0$ in the preceding section. In the absence of $\xi$ or for $\xi=0$ the extremum condition \eqref{UA} for $W(l)$ becomes
\be\label{G1}
\tilde F+2\tilde G\Lambda l^2=0.
\ee
Stability of this extremum requires
\be\label{G2}
\tilde F\geq 0.
\ee
This coincides with the stability condition of positive or zero eigenvalues of the mass matrix \eqref{UC} for solutions with $\xi=0$. In the absence of a dilaton field or for $\xi=0$ the four-dimensional effective Planck Mass reads 
\be\label{G3}
\chi^2=-2\tilde Gl^{-2}.
\ee
It is a positive only for 
\be\label{G4}
\tilde G<0.
\ee
We conclude that stable solutions can exist for positive or vanishing $\Lambda$,
\be\label{G5}
\Lambda=-\frac{\tilde F}{2\tilde G l^2}.
\ee
From this point of view  extrema with nonzero $\Lambda$ are no longer excluded and could be reached for the asymptotic solution.

The difference between solutions in the flat or non-flat phases persists, however. Very large classes of dilatation symmetric $\Gamma$ will still have extrema that lead to $\Lambda=0$. First of all, we recall that  ${\cal M}^4\times T^D$ remains an acceptable stable solution in the flat phase with $\Lambda=0$. The ``size`` of the flat phase depends on the properties of $F$. In this context we recall that $F$ is not arbitrary but corresponds to an ultraviolet fixed point. This may well single out a specific form, for example a simple polynomial.

For large classes of effective actions it is easy to find a very extended space of solutions in the flat phase, including geometries with non-abelian symmetries. As a first example, we may assume that the ultraviolet fixed point corresponds to an effective action where $F$ takes the form 
\be\label{G6}
F=\hat R^\kappa H~,~\kappa>1.
\ee
Here $H$ is an arbitrary (possibly non-linear and non-local) invariant not involving any parameter with dimension of mass and scaling appropriately under dilatations. Obviously, $F$ has an extremum for $\hat R=0$, since 
\be\label{G7}
\delta F=\hat R^{\kappa-1}
(\kappa H\delta \hat R+\hat R\delta H).
\ee
All such solutions corresponds to an extremum of $W$ and we find directly $\tilde F=0, \Lambda=0$. There are many geometries with $\hat R=0$, for example a direct product ${\cal M}^4\times S^{D_1}\times{\cal N}^{D-D_1}$ with ${\cal N}^{D-D_1}a$ space with finite volume and negative curvature scalar which cancels the positive curvature scalar of $S^{D_1}$.

We may replace $S^{D_1}$ by $S^{D_2}\times S^{D_1-D_2}$. Solutions with $\hat R=0$ exist for arbitrary relative radii $r_{D_2}$ and $r_{D_3}$ of the $D_2$ and $D_3=(D_1-D_2)$-dimensional subspaces. The four-dimensional effective potential will therefore not depend on the ratio $\omega=r_{D_2}/r_{D_3}$. We expect a scalar field that corresponds to a variation of $\omega$. In the asymptotic limit $t\to\infty$ the potential for this field vanishes and the scalar becomes massless. We may denote by ``geometrons'' such effective four-dimensional scalars which do not change the value of $F$ at the extremum. They reflect possible deformations of the geometry which are compatible with the extremum value of $F$. In this sense they are somewhat analogous to the moduli fields in string theory. Depending on the particular geometry of the extremum of $F$ in eq. \eqref{G6} there may be several geometrons. In the present cosmological epoch the asymptotic solution is not yet reached and one expects a nontrivial potential for the geometrons which only disappears in the asymptotic limit, e.g. for $l\to 0$. In consequence, the geometrons still have a mass. It is an interesting question if geometrons could play the role of dark matter, similar to axions.

As a second example with a large variety of solutions in the flat phase we consider
\be\label{G8}
F=K^2H,
\ee
with $K$ and $H$ arbitrary scalars built from the metric (without involving couplings with dimension of mass). Again, $F$ has an extremum at $F=0$, this time for $K=0$, since
\be\label{G9}
\delta F=2HK\delta K+K^2\delta H.
\ee
The cosmological constant vanishes again, since $\tilde F=0$. There may be many geometries consistent with the condition $K=0$ and we expect the presence of geometrons for many solutions. The solution persists if we replace $K^2$ by $K^\kappa,\kappa>1$. One realizes that the example \eqref{G6} becomes a special case of this more general class, with $K=\hat R$. If $K$ does not vanish for $\hat R=0$ the solutions in the flat phase will have a non-vanishing curvature scalar. A possible simple example is $K=\hat R_{\hat\mu\hat\nu}\hat R^{\hat\mu\hat\nu}-a\hat R^2$, which can vanish for $\hat R\neq 0$.

What is characteristic for our examples is the existence of an extremum for $W$ for arbitrary dimensionless parameters characterizing $H$ or $K$. Changing these couplings typically changes the details of the geometries which extremize $F$ and $W$. Nevertheless, the four-dimensional cosmological stays zero for all values of such effective couplings. We conclude that a self-tuning mechanism for $\Lambda$ is at work such that $\Lambda$ can readjust to zero whenever these couplings are changed. This is rather different from the situation encountered in four-dimensional theories. We comment on the importance of higher dimensions for the self-tuning of $\Lambda$ in the conclusions. The only thing that is needed for solutions in the flat phase beyond the trivial ${\cal M}^4\times T^D$ geometries is an extremum of $W$ for such solutions. An extremum of $F$ is sufficient for this purpose since extrema of $F$ can occur only for $F_0=0$. This can be shown using the scaling $F\to\alpha^{-d}F$ under the scaling $\hat g\hmn\to\alpha^2\hat g\hmn$, employing arguments analogous to the introduction.

A third example uses a polynomial form
\be\label{G10}
F=F_1-\hat R H
\ee
where $F_1$ contains only terms which are at least quadratic in the Ricci tensor $\hat R\hmn$, while $H$ only involves the totally antisymmetric part $\hat C\hmnrs$ of the curvature tensor. An extremum occurs for
\be\label{G11}
\hat R\hmn=0~,~\hat C\hmnrs\neq 0
\ee
with $\hat C\hmnrs$ chosen such that $H$ is constant, $H_0>0$. Again one has $\tilde F=0,\Lambda=0$. What is less obvious for this extremum is the stability of the solution. For our two first examples we may take $\kappa$ even and $H$ to be positive for the extremum. Then the effective potential after dimensional reduction obeys $V\geq 0$, at least in a region of field space around the extremum, with $V=0$ at the extremum. This guarantees stability with all mass terms for scalars positive are zero. No such simple argument is available for the example \eqref{G10}.

While $\tilde F=0$ is achieved quite easily, we still have to verify $\tilde G<0$ in order to have an acceptable four-dimensional gravity. For the flat geometry ${\cal M}^4\times T^D$ an expansion of $F$ will not contain a term linear in $R^{(4)}$ and therefore $\tilde G=0$. This also holds for an effective action of the type \eqref{G6} where we take $\kappa=2$ for simplicity. With $\hat R=R\nt+R\iv/\sigma$ (cf. eq. \eqref{XS3}) we obtain
\be\label{G12}
\hat R^2=(R\nt)^2+2R\nt R\iv/\sigma+(R\iv)^2/\sigma^2.
\ee
For a minimum of $W$ at $R\nt=0$ the term linear in $R\iv$ vanishes and again $\tilde G=0$. This would lead to an effective gravity with $(R\iv)^2$ - terms and vanishing Planck mass, not compatible with observation. We observe, however, that the vanishing of $\tilde G$ is not a generic feature of all extrema of $W$. For the example \eqref{G10} one finds
\be\label{GB}
\tilde G=-l^2\int_y(g\iD)^{1/2}\sigma H_0<0.
\ee
If the second derivatives of the scalar potential show no instability this type of effective action would generate a reasonable ground state with $\Lambda=0,\chi^2>0$. For progress on these issues more insight about the form of $F$ for an ultraviolet fixed point of gravity would be most welcome.

\section{CONCLUSIONS}
\label{conclusions}

Our detailed investigation of warped geometries in $d$-dimensional gravity with dilatation symmetry has taught us interesting lessons. First of all, not all solutions of the $d$-dimensional field equations correspond to an extremum of the quantum effective action. While for the general solutions of the field equations the four-dimensional cosmological constant $\Lambda$ appears as a free integration constant, the extremum condition for $\Gamma$ typically fixes $\Lambda$. We have given a detailed description of this issue in sect. \ref{potentialfor} in terms of a potential for integration constants.

Second, a dilatation symmetric $\Gamma$ always has extrema for which $\Lambda=0$. This does not depend on the detailed form of $\Gamma$. In particular, $\Gamma$ may be parameterized by a certain number of dimensionless coupling $G_k$. If we vary $G_k$ the extrema in the ``flat phase'' with $\Lambda=0$ persist. This holds even though the detailed geometry of the extremum may vary with $G_k$. We conclude that a self-adjustment mechanism for $\Lambda$ is at work, which always permits $\Lambda$ to adapt to zero even for a variation for internal geometry.

The existence of more than four dimensions is crucial for the possibility of this self-tuning. This can be understood by a simple comparison. In a four-dimensional theory we typically have a finite number of scalar fields that we may denote by $\alpha_i$, and a number of couplings $G_k$. The effective potential $V$ is a function of $\alpha_i$, depending on the couplings $G_k$. An extremum of the effective action with constant scalar fields and $\Lambda=0$ requires 
\be\label{CC1}
\frac{\partial V}{\partial\alpha_i}(\alpha_i;G_k)|_{\alpha^{(0)}_i}=0~,~V(\alpha_i^{(0)};G_k)=0.
\ee
Even though it may be possible that eqs. \eqref{CC1} are obeyed for a certain choice of $G_k$, this typically no longer holds for neighboring $G_k$. Thus the couplings $G_k$ have to be tuned to special values for obtaining $\Lambda=0$. Understanding why the couplings take these special values constitutes the cosmological constant problem.

In a higher dimensional settling the scalars $\alpha_i$ are replaced by functions of the internal coordinates $y$. The extremum conditions for $V$ are now replaced by an extremum condition for a functional $W[\alpha_i(y); G_k]$, i.e.
\be\label{CC2}
\delta W [\alpha_i(y);G_k]=0.
\ee
This extremum condition typically has solutions. In the case of regular geometries eq. \eqref{CC2} is equivalent to the $D$-dimensional field equations, while additional extremum constraints restrict the space of solutions in case of singularities. For a dilatation symmetric $\Gamma$ we have shown that eq. \eqref{CC2} implies $W=0$ for the extremum, and in consequence $\Lambda=0$.

The role of the additional internal dimensions is twofold. First, the flexibility of the adjustment of functions $\alpha_i(y)$ guarantees that an extremum of $W$ persists if we vary the couplings $G_k$. These functions may be viewed as infinitely many four-dimensional scalar fields. Second, in case of dilatation symmetry the scaling \eqref{2F} guarantees that any extremum of $W$ occurs at $W_0=0$, implying $\Lambda=0$. For obtaining $\Lambda=0$ for a given parameter set $\{G_k\}$ it is therefore sufficient to find an extremum of $W$. 

In summary, we can state a simple condition for finding an extremum of the dilatation symmetric $d$-dimensional effective action $\Gamma$ which leads, after dimensional reduction, to a vanishing four-dimensional cosmological constant $\Lambda=0$: the functional $W$ should have an extremum. 

This condition amounts to the existence of an extremum of the action for a particular $D$-dimensional euclidean gravity theory for the metric and scalar fields. We may demonstrate this in the absence of a higher-dimensional dilaton field $\xi$. Then 
\be\label{C1}
W=\int_y \bar g^{1/2}\sigma^{\frac d2}F[\hat g\hmn]
\ee
is evaluated for 
\be\label{C2}
\hat g\hmn=\sigma(y)\bar g\hmn~,~\bar g\hmn =
\left(\begin{array}{ccc}
\eta_{\mu\nu}&,&0\\0&,&\bar g_{\alpha\beta}
\end{array}\right),
\ee
where we realize the connection to the ansatz \eqref{2c} with $g\iv_{\mu\nu}=\eta_{\mu\nu}$ and $g\iD_{\alpha\beta}=\bar g_{\alpha\beta}\sigma$. Inserting eq. \eqref{C2} is equivalent to a $D$-dimensional Weyl scaling and results in
\be\label{C3}
W=\int_y\bar g^{1/2}
\{\bar F[\bar g_{\alpha\beta}]+
\bar K[\sigma,\bar g_{\alpha\beta}]\}.
\ee
Here $\bar F$ obtains from $F$ by replacing $\hat g\hmn\to\bar g\hmn$,
\be\label{C4}
\bar F[\bar g_{\alpha\beta}]=F[\bar g\hmn],
\ee
and exploiting the direct product structure of $\bar g\hmn$ \eqref{C2}.The kinetic term $\bar K$ contains derivatives $\partial_\alpha\sigma$ and vanishes for constant $\sigma$. Both $\bar F$ and $\bar K$ are scalars with respect to $D$-dimensional general coordinate transformations and scale under constant rescalings of $\bar g_{\alpha\beta}$ as
\be\label{C5}
\bar g_{\alpha\beta}\to\alpha^2\bar g_{\alpha\beta}~,~
\sigma\to\sigma\Rightarrow\bar F\to\alpha^{-d} F~,~\bar K\to \alpha^{-d}\bar K,
\ee
such that eq. \eqref{2F} holds, $W\to\alpha^{-4}W$. The inclusion of a $d$-dimensional dilaton is straightforward - the functional $W$ in eq. \eqref{C3} contains then additional terms involving an additional scalar field $\xi(y)$. 

The functional $W$ always has extrema. Thus $\Gamma$ always has extrema in the flat phase with $\Lambda=0$. The interesting issue is not anymore to find extrema with $\Lambda=0$, but rather to find interesting extrema beyond the trivial case where $\bar g_{\alpha\beta}$ is the flat metric of a torus $T^D$ and $\sigma$ is constant. Furthermore, the physically interesting extrema should result after dimensional reduction in a nonvanishing and finite Planck mass $\chi$. If the asymptotic behavior of a cosmological runaway solution approaches a region in field space where $\Gamma$ becomes dilatation symmetric the cosmological constant problem can be solved. The issue remains then to find an interesting particle physics. 

The effective four-dimensional theory obtained by dimensional reduction from a dilatation symmetric $\Gamma$ always contains a massless scalar field, the dilaton. For pure higher dimensional gravity the dilaton can by associated with a variation of the characteristic length $l$ of internal space. If a cosmological runaway solution approaches the region in field space where $\Gamma$ becomes dilatation symmetric, but has not yet reached the fixed point, some residual ``dilatation anomaly'' remains. This generates an effective potential for the pseudo-dilaton or cosmon. After Weyl scaling this potential V($\varphi$) has to vanish as the value of the cosmon $\varphi$ moves to infinity. (We use conventions where dilatation symmetry is realized for $\varphi \to \infty$.) We can associate $V (\varphi)$, together with the kinetic energy of the cosmon, with dark energy. Our setting realizes a quintessence scenario where dark energy relaxes to zero for $t\to\infty$, rather than approaching a nonzero cosmological constant. The huge age of the universe in units of the Planck time can explain why the dark energy density today is tiny in units of the Planck mass \cite{CWQ}.

The extremum of $W$ can be degenerate geometrically in the sense that families of extrema exist which are parameterized by continuous dimensionless parameters $\omega_j$ beyond the degeneracy in the characteristic length scale $l$. For a dilatation symmetric $\Gamma$ the effective four-dimensional actions exhibits in this case massless scalars beyond the dilaton. We call these degrees of freedom ``geometrons'' since they typically reflect a change of internal geometry at fixed $l$. (In the presence of a higher dimensional scalar field $\xi$ one of the $\omega_j$ may also be related to the dimensional ratio $\bar \xi l ^{\frac{d-2}{2}}$, without influencing the ``shape'' of internal geometry). The role of such geometrons for the cosmology in the present epoch depends on the details of the dilatation anomaly, since the latter will be responsible for providing a potential and mass terms for the $\omega_j$. If one or several masses for geometrons are in the appropriate range the geometrons may constitute interesting candidates for dark matter.

We finally mention that the presence of an ultraviolet fixed point for gravity, which is approached in field space for $l \to 0$, could also be helpful for another big problem for the unification of gravity and gauge interactions in the context of higher-dimensional theories. This problem is the lack of knowledge in the choice of the higher-dimensional action. Since the predictions for the particle content and interactions of the effective four-dimensional theory depend on the details of the higher-dimensional effective action, this type of unification has so far suffered from a lack of predictivity for experiments. If the cosmological evolution drives the fields to a region which is dominated by an ultraviolet fixed point, the form of $\Gamma$ is no longer arbitrary. Typically the fixed point itself may not involve any free parameters and have a unique form of $\Gamma$ for a given higher-dimensional field content. Only a few relevant (or marginal) parameters may describe the deviations from the fixed point. Predictions for experiments in the present universe will then only depend on these few relevant parameters. Needless to say that the great challenge remains to establish such an ultraviolet fixed point of higher-dimensional gravity.

\section*{APPENDIX A: Warped branes and five dimensional dilatation symmetry}
\label{appendixA}
\renewcommand{\theequation}{A.\arabic{equation}}
\setcounter{equation}{0}
The case $D=1$ is special. The metric is now only characterized by the function $\sigma(z)$ and one has the field equations \eqref{18}, \eqref{19}, \eqref{21}
\begin{eqnarray}\label{AA1}
&&\delta'=E\sigma^{-2},\\
&&-2\Lambda\sigma^{-1}+3\frac{\sigma''}{\sigma}=-\delta'^2,\label{AA2}\\
&&-4\Lambda\sigma^{-1}+3\left(\frac{\sigma'}{\sigma}\right)^2=\delta'^2.\label{AA3}
\end{eqnarray}
They can be combined to a second order differential equation
\begin{equation}\label{AA4}
\frac{\sigma''}{\sigma}+\left(\frac{\sigma'}{\sigma}\right)^2=2\Lambda\sigma^{-1},
\end{equation}
or, for $s=\ln(\sigma/\sigma_0)$, 
\begin{equation}\label{AA5}
s''+2s'^2-2\tilde\Lambda e^{-s}=0,
\end{equation}
which is equivalent to eq. \eqref{31B}. For $\tilde \Lambda=0$ the first order differential equation for $U=s'=\sigma'/\sigma$, 
\begin{equation}\label{AA6}
U'=-2U^2,
\end{equation}
has the general solution (with integration constant absorbed in the location of the singularity
\begin{equation}\label{AA7}
U=\frac{1}{2z},~s=\frac12\ln z,~\sigma=\sigma_0 z^{\frac12}.
\end{equation}
This solution does not lead to acceptable four-dimensional gravity since $\int_z\sigma^2$ diverges for $z\to\infty$.

For $\tilde\Lambda\neq 0$ the importance of the term $\sim \tilde\Lambda$ may be estimated by inserting $s=\ln z/2$, 
\begin{equation}\label{AA8}
\tilde \Lambda e^{-s}=\tilde\Lambda z^{-1/2}.
\end{equation}
It will become important as $z$ increases, since $s''$ and $s'^2$ decrease faster $\sim z^{-2}$. For $\tilde\Lambda\neq 0$ eq. \eqref{AA5} describes the damped motion of a particle in a potential 
\begin{equation}\label{AA9}
V=2\tilde\Lambda e^{-s}.
\end{equation}
For $\tilde\Lambda>0$ the potential decreases for $s\to\infty$, and one finds the particular solution
\begin{equation}\label{AA10}
s=2\ln z+\ln(\tilde\Lambda/3).
\end{equation}
In terms of $\sigma(z)$ this reads
\begin{equation}\label{AA11}
\sigma=\frac{\Lambda}{3}z^2.
\end{equation}
No four-dimensional gravity exists since $\int_z\sigma^2$ is divergent. The general solution approaches the asymptotic solution \eqref{AA11} for $z\to\infty$, for example by switching from the singular solution \eqref{AA7} to \eqref{AA11} at $z_c\approx(\tilde\Lambda/3)^{-2/3}$. 

On the other hand, for negative $\tilde\Lambda$ the potential increases for $s\to\infty$. The increase of $s$ is stopped at a turning point, where $s''(z_t)=2\tilde\Lambda e^{-s(z_t)}<0$. After reaching its maximum value at the turning point $z_t$, the function $\sigma(z)$ decreases again until it reaches zero at a second singularity $\bar z$. The solutions with two singularities at $z=0$ and $z=\bar z$ will lead to a finite four-dimensional Planck mass. If $|\Lambda|$ is small enough, this yields an acceptable four-dimensional gravity. 

The precise solution depends on the integration constants $\Lambda,~\sigma(z_0)$ and $\sigma'(z_0)$, with $z_0$ a suitable initial value for $z$. Eq. \eqref{AA2} or \eqref{AA3} relates these integration constants with $E$. In fact, the general solution of eq. \eqref{AA4} has no information about the scalar field - the latter enters via the integration constants. We may use an alternative starting point for the solution of the system of equations \eqref{AA1}-\eqref{AA3} inserting eq. \eqref{AA1} into eq. \eqref{AA2}
\begin{equation}\label{AA12}
\sigma''-\frac{2}{3}\Lambda+\frac{E^2}{3}\sigma^{-3}=0.
\end{equation}
Now $\sigma(z)$ is the analogue of the undamped motion of a particle in a potential
\begin{equation}\label{AA13}
\tilde V=-\frac23\Lambda\sigma-\frac{E^2}{6}\sigma^{-2},
\end{equation}
such that the ``kinetic energy'' obeys
\begin{equation}\label{AA14}
\sigma'^2=2(\epsilon-\tilde V).
\end{equation}
The integration constant $\epsilon$ is fixed by eq.\eqref{AA3} as $\epsilon=0$, i.e.
\begin{equation}\label{AA15}
\sigma'^2=\frac{E^2}{3\sigma^2}+\frac{4\Lambda\sigma}{3}.
\end{equation}

For positive $\Lambda$ and large $\sigma$ we recover eq. \eqref{AA11},
\begin{equation}\label{AA16}
\sigma'^2=\frac49\Lambda^2z^2=\frac{4\Lambda\sigma}{3},
\end{equation}
while for small $\sigma$ we find eq. \eqref{AA7} with
\begin{equation}\label{AA17}
\frac{E^2}{\sigma^4_0}=\tilde E^2=\frac34.
\end{equation}
The latter is the exact solution for $\Lambda=0$. For $\Lambda<0$ the maximal value of $\sigma$ is given by 
\begin{equation}\label{AA18}
\sigma(z_t)=\left(-\frac{E^2}{4\Lambda}\right)^{1/3}.
\end{equation}
The solution is symmetric around the turning point, which is therefore in the middle between the singularities at $z=0$ and $z=\bar z$, i.e. $z_t=\bar z/2$. Sufficiently away from the turning, point we find again the singular behavior \eqref{AA7}, \eqref{AA17},
\begin{equation}\label{AA19}
\sigma(z\to 0)=\sigma_0 z^{1/2},~\sigma(z\to\bar z)=\sigma_0(\bar z-z)^{1/2}.
\end{equation}
A qualitatively correct approximation is 
\begin{equation}\label{AA20}
\sigma=\sigma_0\bar z^{-1/2}~z^{1/2}(\bar z-z)^{1/2},
\end{equation}
with an approximate turning point
\begin{eqnarray}\label{AA21}
\sigma(z_t)&=&\sigma(\bar z/2)=\frac{\sigma_0}{2}\bar z^{1/2}~,\nonumber\\
\bar z&=&\left(-\frac{2E^2}{\Lambda\sigma^3_0}\right)^{2/3}=
\left(-\frac{2\tilde E^2}{\tilde \Lambda}\right)^{2/3}.
\end{eqnarray}

\section*{APPENDIX B: Hyperbolic Einstein spaces with isometries and finite volume}
\renewcommand{\theequation}{B.\arabic{equation}}
\setcounter{equation}{0}
In this appendix we study Einstein spaces in $E$ dimensions,
\begin{equation}\label{B1}
\bar R_{\bar\alpha\bar\beta}=C g_{\bar\alpha\bar\beta}.
\end{equation}
The signature of the metric is euclidean. We are particularly interested in hyperbolic spaces where $C$ is negative. If they have a finite volume, such spaces will be candidates for the discussion in sect. \ref{warped}, with $E=D-1$. The maximal isometry for an $E$-dimensional space, $SO(E+1)$, implies positive $C=E-1$. We therefore consider a smaller isometry group $SO(D_1+1)\times SO(D_2+1)$, with $D_1+D_2+1=E$. The most general metric  can be written in a form similar to the ansatz \eqref{4}
\begin{equation}\label{B2}
\bar g_{\bar\alpha\bar\beta}(x,y,z)=
\left(
\begin{array}{ccccc}
\sigma(z)g^{(D_1)}_{\mu\nu}(x)&,&0&,&0\\
0&,&\rho(z)g^{(D_2)}_{\alpha\beta}(y)&,&0\\
0&,&0&,&1
\end{array}
\right).
\end{equation}
The coordinates $x^\mu,\mu=1\dots D_1$, describe a $D_1$-dimensional Einstein space, and similarly the coordinates $y^\alpha, \alpha=1\dots D_2$, parameterize a $D_2$-dimensional Einstein space, with Ricci tensors
\begin{equation}\label{B3}
R^{(D_1)}_{\mu\nu}=\Lambda_1 g^{(D_1)}_{\mu\nu}~,~R^{(D_2)}_{\alpha\beta}=\Lambda_2 g^{(D_2)}_{\alpha\beta}.
\end{equation}
For a realization of the isometries $SO(D_1+1)\times SO(D_2+1)$ the Einstein spaces are spheres, with $\Lambda_1=D_1-1~,~\Lambda_2=D_2-1$. We will keep the discussion more general at this stage. (For $D_1=1~,~D_2=9$ the isometry group is $SO(10)\times SO(2)$. This could account in the dimensionally reduced theory discussed in sect. IV for a realistic grand unified gauge group $SO(10)$ and a generation symmetry $SO(2)$.) Our notation is somewhat similar to sect. IV in order to display the analogy for $D_1=4$, where only the signature differs. The reader should not get confused by this, the coordinates $x^\mu$ have no relation to the space-time coordinates in sect. IV. Also the $z$ and $y$ coordinates have a different meaning.

The field equation \eqref{B1} for this metric can be taken from \cite{RDW}. (We employ here a different signature for the metric.) They read 
\begin{eqnarray}\label{B4}
&&(D_1+D_2-1)C-(D_1-2)\Lambda_1\sigma^{-1}-D_2\Lambda_2\rho^{-1}\nonumber\\
&&+(D_1-1)\frac{\sigma''}{\sigma}+\frac14(D_1-1)(D_1-4)\frac{\sigma'^2}{\sigma^2}\nonumber\\
&&+\frac12(D_1-1)D_2\frac{\rho'}{\rho}\frac{\sigma'}{\sigma}\nonumber\\
&&+D_2\frac{\rho''}{\rho}+\frac14D_2(D_2-3)\frac{\rho'^2}{\rho^2}=0,\\
&&(D_1+D_2-1)C-D_1\Lambda_1\sigma^{-1}-(D_2-2)\Lambda_2\rho^{-1}\label{B5}\nonumber\\
&&+(D_2-1)\frac{\rho''}{\rho}+\frac14
(D_2-1)(D_2-4)\frac{\rho'^2}{\rho^2}\nonumber\\
&&+\frac12(D_2-1)D_1\frac{\rho'}{\rho}\frac{\sigma'}{\sigma}\nonumber\\
&&+D_1\frac{\sigma''}{\sigma}+\frac14D_1(D_1-3)\frac{\sigma'^2}{\sigma^2}=0,\\
&&(D_1+D_2-1)C-D_1\Lambda_1\sigma^{-1}-D_2\Lambda_2\rho^{-1}\nonumber\\
&&+\frac12 D_1D_2\frac{\rho'}{\rho}\frac{\sigma'}{\sigma}+\frac14 D_1(D_1-1)\frac{\sigma'^2}{\sigma^2}\label{B6}\nonumber\\
&&+\frac14 D_2(D_2-1)\frac{\rho'^2}{\rho^2}=0.
\end{eqnarray}
The discussion of possible solutions can be done in parallel with sects. \ref{warped}-\ref{global}

An interesting special case are two-dimensional spaces, $E=2$, that we may obtain for $D_1=1,D_2=0$. In this case eq. \eqref{B5} and all terms involving $\rho'$ and $\rho''$ in eqs. \eqref{B4}, \eqref{B6} have to be omitted. Since $\Lambda_1=\Lambda_2=0$ eqs. \eqref{B4} and \eqref{B6} are obeyed identically for this case. In fact, these equations amount to the identity
\begin{equation}\label{B6A}
H_{\hat\mu\hat\nu} =R_{\hat\mu\hat\nu} -\frac12 R g_{\hat\mu\hat\nu} =0,
\end{equation}
that is valid for arbitrary $C$. Since for $E=2$ the Einstein tensor is traceless and vanishes identically for our ansatz, we need in this case an additional equation for the curvature scalar
\begin{equation}\label{B6B}
R=2C=-\frac{\sigma''}{\sigma}+\frac12\left(\frac{\sigma'}{\sigma}\right)^2.
\end{equation}
In terms of $U$ this reads
\begin{equation}\label{B6C}
U'+\frac12 U^2+2C=0.
\end{equation}
For $C>0$ this has the general solution
\begin{equation}\label{B6D}
\sigma=\frac{\sigma_0}{C}\sin^2(\sqrt{C}z).
\end{equation}
For $\sigma_0=1$ this describes a sphere, while for $\sigma_0\neq 1$ the geometry is singular for $z=0$ and $z=\pi/\sqrt{C}$. For $\sigma_0<1$ one encounters conical singularities, with deficit angle
\begin{equation}\label{B6E}
\Delta=2\pi(1-\sqrt{\sigma_0}).
\end{equation}
This geometry has the shape of an American football.

For $C<0$ we may infer the general solution by analytic continuation
\begin{equation}\label{B6F}
\sigma=\frac{\sigma_0}{|C|}\sinh^2(\sqrt{|C|}z).
\end{equation}
This geometry has again a canonical singularity for $z=0$ if $\sigma_0<1$, and is regular for $z=0$ for $\sigma_0=1$. However, $\sigma$ diverges $\sim\exp(2\sqrt{|C|}{|z|})$ for $|z|\to\infty$. The resulting geometry does not have a finite volume anymore - the integral $\int_z\sigma^{1/2}(z)$ diverges for $z\to\infty$. We conclude that in two dimensions negatively curved Einstein spaces with continuous isometries and finite volume do not exist. This shows that the possible existence for $E>2$ of geometries obeying the ansatz \eqref{B2} and with finite volume is not trivial.

We now turn to $D_1\geq 1,~D_2\geq 1$ for which the three equations \eqref{B4}-\eqref{B6} have to be obeyed. In particular, we find the relation
\begin{eqnarray}\label{B7}
&&\frac{\rho'}{\rho}\frac{\sigma'}{\sigma}=-\frac{4C}{D_2}+\frac{4}{D_2}
\frac{\Lambda_1}{\sigma}-\frac{2}{D_2}
\frac{\sigma''}{\sigma}-\frac{D_1-2}{D_2}\frac{\sigma'^2}{\sigma^2}\nonumber\\
&&=-\frac{4C}{D_1}+\frac{4}{D_1}\frac{\Lambda_2}{\rho}-\frac{2}{D_1}\frac{\rho''}{\rho}
-\frac{D_2-2}{D_1}\frac{\rho'^2}{\rho^2}.
\end{eqnarray}
The structure of possible singularity is similar to sect. IV, 
\begin{eqnarray}\label{B8}
\sigma&=&\sigma_0 z^{-\eta}~,~\rho=\rho_0 z^\gamma~,~\gamma=\frac{2+D_1\eta}{D_2},\nonumber\\
\eta&=&\frac{-2\pm 2\sqrt{D_2(D_1+D_2-1)/D_1}}{D_1+D_2}
\end{eqnarray}
and similar for $z\to\bar z$. The volume of the $E$-dimensional space reads
\begin{equation}\label{B8A}
\Omega_E=\Omega_{D_1}\Omega_{D_2}\tilde\Omega_E~,~\tilde\Omega_E\int_z\sigma^{\frac{D_1}{2}}
\rho^{\frac{D_2}{2}},
\end{equation}
with $\Omega_{D_1},\Omega_{D_2}$ the volumes of the subspace with metric $g^{(D_1)}_{\mu\nu}$ and $g^{(D_2)}_{\alpha\beta}$. For a singularity \eqref{B8} at $z=0$ the relation between $\gamma$ and $\eta$ implies $\tilde\Omega_E\sim\int_z z$. Geometries with two singularities at $z=0,\bar z$ have a finite volume.

for an investigation if a given solution corresponds to a finite volume $\Omega_E$ it is useful to define the function
\begin{equation}\label{B8B}
v(z)=\sigma(z)^{\frac{D_2}{2}}\rho(z)^{\frac{D_2}{2}}.
\end{equation}
For all values of $z$ where our coordinates are well defined $v(z)$ must be strictly positive. If this range of $z$ extends to $z\to\infty$, a finite volume requires that $v$ approaches zero faster than $z^{-1}$, and similar for $z\to-\infty$. If the range of $z$ is bounded by a singularity \eqref{B8}, we know that $v$ vanishes $\sim z$ or $\sim(\bar z-z)$. Finally, if $z=0$ corresponds to a regular point where $\sigma$ is constant and $\rho$ vanishes $\sim z^2$, we also find a vanishing $v$. For any geometry with finite volume $v(z)$ vanishes at the boundaries of the allowed range of $z$ and is positive inside this range. We conclude that a finite volume requires the existence of a maximum of $v(z)$, or a point $z_m$ where 
\begin{eqnarray}\label{B8C}
&&D_1\frac{\sigma'}{\sigma}+D_2\frac{\rho'}{\rho}=0\\
&&D_1\left(\frac{\sigma''}{\sigma}-\left(\frac{\sigma'}{\sigma}\right)^2\right)+D_2
\left(\frac{\rho''}{\rho}-\left(\frac{\rho'}{\rho}\right)^2\right)<0.\nonumber
\end{eqnarray}

From eq. \eqref{B6} we infer that $\sigma'/\sigma$ must obey for $z=z_m$
\begin{eqnarray}\label{B8D}
&&\frac{D_1(D_1+D_2)}{4D_2}\left(\frac{\sigma'}{\sigma}\right)^2\\
&&=(D_1+D_2-1)C-\frac{D_1\Lambda_1}{\sigma}-\frac{D_2\Lambda_2}{\rho}=A.\nonumber
\end{eqnarray}
This is possible only if $A>0$, and a finite volume therefore requires the condition
\begin{equation}\label{B8E}
\frac{D_1\Lambda_1}{\sigma}+\frac{D_2\Lambda_2}{\rho}<(D_1+D_2-1)C.
\end{equation}
For negative $C$ this inequality can be obeyed only if $\Lambda_1$ or $\Lambda_2$ are negative. We conclude that hyperbolic Einstein spaces with the ansatz \eqref{B2} and positive or vanishing $\Lambda_1,\Lambda_2$ have always infinite volume, similar to the geometry \eqref{B6F}.

If either $\Lambda_1$ or $\Lambda_2$ are negative, the condition \eqref{B8E} can always be obeyed by choosing a small enough $\sigma(z_m)$ or $\rho(z_m)$. The second condition in \eqref{B8C} reads then
\begin{equation}\label{B8F}
D_1\frac{\sigma''}{\sigma}+D_2\frac{\rho''}{\rho}<4A.
\end{equation}
Eqs. \eqref{B4}, \eqref{B5} yield from $z_m$
\begin{eqnarray}\label{B86}
&&(D_1-1)\frac{\sigma''}{\sigma}+D_2\frac{\rho''}{\rho}=2A
\left(1-\frac{2D_2}{D_1(D_1+D_2)}\right)-2\frac{\Lambda_1}{\sigma},\nonumber\\
&&D_1\frac{\sigma''}{\sigma}+(D_2-1)\frac{\rho''}{\rho}=2A
\left(1-\frac{2D_1}{D_2(D_1+D_2}\right)-2\frac{\Lambda_2}{\rho}.\nonumber\\
\end{eqnarray}
and therefore
\begin{eqnarray}\label{B8H}
&&\frac{\sigma''}{\sigma}=2A
\left\{\frac{2D_2}{D_1(D_1+D_2)}-\frac{1}{D_1+D_2-1}\right\}\nonumber\\
&&+\frac{2(D_2-1)}{D_1+D_2-1}\frac{\Lambda_1}{\sigma}-\frac{2D_2}{D_1+D_2-1}\frac{\Lambda_2}{\rho},\nonumber\\
&&\frac{\rho''}{\rho}=2A\left\{\frac{2D_1}{D_2(D_1+D_2)}-\frac{1}{D_1+D_2-1}\right\}\nonumber\\
&&+\frac{2(D_1-1)}{D_1+D_2-1}\frac{\Lambda_2}{\rho}-\frac{2D_1}{D_1+D_2-1}\frac{\Lambda_1}{\sigma}.
\end{eqnarray}
The condition \eqref{B8F} becomes
\begin{equation}\label{B8I}
(D_1+D_2-2)\left(\frac{D_1\Lambda_1}{\sigma}+\frac{D_2\Lambda_2}{\rho}\right)<
(D_1+D_2)(D_1+D_2-1)C
\end{equation}
and can be fulfilled for either $D_1$ or $D_2$ larger than one and small enough $\sigma(z_m)$ or $\rho(z_m)$. We can solve numerically the field equations, starting at $z_m$ with eq. \eqref{B8D}, and we find indeed solutions with finite volume $\Omega_E$. 

If the conditions $A>0$ and \eqref{B8I} cannot be met the volume $\Omega_E$ is necessarily infinite. One can understand this issue from a different perspective. Let us concentrate, for simplicity, on $D_1=1$ where $\Lambda_1=0$. With
\begin{equation}\label{B9}
U=\sigma'/\sigma~,~W=\frac{\rho'}{\rho}=-\frac{4C}{D_2U}-\frac{1}{D_2}
\left(2\frac{U'}{U}+U\right)
\end{equation}
one finds a second order differential equation for $U$
\begin{eqnarray}\label{B10}
&&D_2U''-(D_2+1)\frac{U'^2}{U}-2(D_2+2)C\frac{U'}{U}\nonumber\\
&&\qquad -U'U+\frac{\partial\hat V}{\partial U}=0,\\
&&\frac{\partial\hat V}{\partial U}=-4C^2 U^{-1}-(D_2+2)CU-
\frac{D_2+1}{4}U^3.\nonumber
\end{eqnarray}
This describes the motion of a particle in a potential 
\begin{equation}\label{B11}
\hat V(U)=-4C^2\ln U-\frac{D_2+2}{2}CU^2-\frac{D_2+1}{16}U^4,
\end{equation}
with damping or antidamping. We note that $\partial\hat V/\partial U$ is negative everywhere if $C$ is positive. For $C<0$ the potential $\hat V(U)$ has a minimum and a maximum for
\begin{equation}\label{B12}
U^2_{min}=-\frac{4C}{D_2+1}~,~U^2_{max}=-4C.
\end{equation}
The possible singularities \eqref{B8} correspond to
\begin{equation}\label{B13}
U=-\frac\eta z~,~W=\frac{\gamma}{z}
\end{equation}
and therefore to large positive or negative $U\to\pm \infty$. 

Solutions for which $U$ approaches $U_{min}$ asymptotically for $z\to\infty$ correspond to a space with infinite volume $\Omega_E$. Indeed, $U=U_{min}=2\sqrt{|C|}/\sqrt{D_2+1}$ (for $C<0$) solves eq. \eqref{B10} and corresponds to an exponentially diverging $\sigma$
\begin{equation}\label{B13A}
\sigma=\bar\sigma\exp(U_{min} z).
\end{equation}
For $D_2=0$ this is the solution \eqref{B6F}. For $D_2\geq 1$ eq. \eqref{B9} implies
\begin{equation}\label{B13B}
W=\frac{\rho'}{\rho}=-\frac{1}{D_2 U_{min}}(4C+U^2_{min})=U_{min},
\end{equation}
such that $\rho$ also diverges exponentially.

In order to see for which initial conditions a solution can approach $U_{min}$, it is instructive to investigate the possible turning points for $U$. For such a turning point at $z_t$ one has 
\begin{equation}\label{B13C}
U'(z_t)=0~,~U''(z_t)=-(1/D_2)\partial\hat V/\partial U.
\end{equation}
Denoting $U_t=U(z_t)~,~W_t=W(z_t)$ one finds
\begin{equation}\label{B14}
W_t=-\frac{4C}{D_2U_t}-\frac{U_t}{D_2},
\end{equation}
and insertion into eq. \eqref{B6} yields the condition
\begin{eqnarray}\label{B15}
&&\frac{D^2\Lambda_2}{\rho_t}=-\frac{D_2+1}{4}U^2_t+
(D^2_2-2)C+4(D_2-1)\frac{C^2}{U^2_t}\nonumber\\
&=&U^{-1}_t\frac{\partial\hat V}{\partial U}(U_t)+D_2(D_2+1)C+4D_2\frac{C^2}{U^2_t}.
\end{eqnarray}
For $\Lambda_2>0$ the r.h.s must be positive and this requires
\begin{eqnarray}\label{B16}
&&U^2_t<-\frac{4C}{D_2+1}=U^2_{min} \quad \textup{for }C<0,\nonumber\\
&&U^2_t<4C(D_2-1) \quad \textup{for }C>0.
\end{eqnarray}
For $\Lambda_2<0$ the opposite inequalities hold.

Solutions with a turning point can be formulated as an initial value problem for the differential equation \eqref{B10}. The initial values are taken at $z_t$, namely $U_t$ and $U'(z_t)=0$, with $U_t$ obeying the condition \eqref{B16} and $\rho_t$ given by eq. \eqref{B15}. For $U_t>0$ the function $U(z)$ takes its minimal value at the turning point. We can take $U>0$ without loss of generality. 

Let us consider $C<0$ and $U_t$ close to the minimum, such that we can linearize in $\delta U=U-U_{min}$, 
\begin{eqnarray}\label{B17}
&&\delta U''+\Gamma m\delta U'+m^2\delta U=0,\nonumber\\
&&m^2=-2C~,~\Gamma=\frac{D_2+3}{\sqrt{2D_2+2}}.
\end{eqnarray}
Since $\Gamma>2$ for all $D_2>1$ the approach to $U_{min}$ is overdamped,
\begin{equation}\label{B18}
\delta U=c_1e^{-\kappa_1mz}+c_2e^{-\kappa_2mz},
\end{equation}
with $\kappa_{1,2}$ obeying $\kappa^2-\Gamma\kappa+1=0$. From eq. \eqref{B9} we infer the asymptotic behavior for $z\to\infty$
\begin{equation}\label{B19}
U=U_{min}~,~\sigma=\bar\sigma e^{U_{min}z}~,~\rho=\bar\rho e^{U_{min}z}
\end{equation}
(For $\Lambda_2=0$ eq. \eqref{B19} becomes an exact special solution for all $z$.) By a numerical solution we find that the asymptotic behavior \eqref{B19} is actually realized for arbitrary $U_t\leq U_{min}$. The situation for $D_1>1$ and $\Lambda_1\geq 0$ is similar. 

For a more general discussion it is useful to simplify the terms involving first derivatives by the choice of new variables (for $U>0$)
\begin{equation}\label{B20}
y=U^{-1/D_2}.
\end{equation}
Eq. \eqref{B20} then reads
\begin{eqnarray}\label{B21}
&&y''+K(y)y'+\frac{\partial\tilde V}{\partial y}=0,\nonumber\\
&&\frac{\partial\tilde V}{\partial y}=\frac{4C^2}{D^2_2}y^{2D_2+1}
+\frac{D_2+2}{D^2_2}Cy+\frac{D_2+1}{4D^2_2}y^{-2D_2+1},\nonumber\\
&&K(y)=-\frac{2(D_2+2)}{D_2}Cy^{D_2}-\frac{1}{D_2}y^{-D_2}.
\end{eqnarray}
(For $U<0$ we take $y=-(-U)^{-1/D_2}$. This changes the sign of $K(y)$.) Consider first $C<0$, where the possible turning point $y_t$ occurs for
\begin{equation}\label{B40A}
y_t>y_{min}=\big[-(D_2+1)/4C\big]^{\frac{1}{2D_2}}.
\end{equation}

The damping coefficient $K(y)$ is positive for large $y>y_d$,
\begin{eqnarray}\label{B22}
y_d&=&\big[-2(D_2+2)C\big]^{-\frac{1}{2D_2}}=\left(\frac{2}{D_2+2}\right)^{\frac{1}{2D_2}}y_{max}
\nonumber\\
&=&\left(\frac{2}{(D_2+2)(D_2+1)}\right)^{\frac{1}{2D_2}}y_{min}.
\end{eqnarray}
For the initial values $y(z_t)=y_t~,~y'(z_t)=0$ one finds $y$ decreasing for $z>z_t$, first in a damped motion. There are two alternatives. Either $y$ decreases beyond $y_{max}$. Then nothing can prevent a further decrease to the singularity for $y\to 0$, with damping turned to antidamping in the vicinity of the singularity. Or $y$ gets sufficiently damped that it cannot reach $y_{max}$. For large $z$  the motion will then end at $y_{min},~y(z\to\infty)\to y_{min}$. For $y_t$ close enough to $y_{min}$ one has $\tilde V(y_t)<\tilde V(y_{max})$, such that a damped motion can only end in the minimum. We recover the overdamped motion \eqref{B18}, \eqref{B19}. We find that for $\Lambda_{1,2}\geq 0~,~C<0$ the solutions with a turning point for $U$ show always the asymptotic behavior \eqref{B19} and do not have finite volume. 

Finally, it is interesting to ask if there are possible solutions where $\sigma$ is not monotonic. In this case $U$ changes sign such that $U(z_e)=0$ at an extremum of $\sigma$ at $z_e$. We note that eq. \eqref{B9} is only valid for $U\neq 0$, while for $U=0$ eq. \eqref{B6} becomes for $D_1=1$
\begin{equation}\label{B23}
\frac14(D_2-1)W^2(z_e)=-C+\frac{\Lambda_2}{\rho(z_e)},
\end{equation}
such that $W$ remains finite. In the vicinity of $z_e$ one can approximate
\begin{eqnarray}\label{B24}
U'&=&-2C~,~U=-2C(z-z_e),\nonumber\\
\sigma&=&\sigma(z_e)\exp \big[-C(z-z_e)^2\big].
\end{eqnarray}
For $C<0$ this describes a minimum of $\sigma$, while for $C>0$ the extremum is a maximum. For $C>0$ such a maximum of $\sigma$ can only occur for $\Lambda_2>0$. A more extensive discussion of the solutions with $C>0$ can be found in ref. \cite{CWCC}, sect. IV, with the identification $2\tilde V=(D_1+D_2-1)C$, and for solutions with $C=0$ in the appendix of \cite{CWA}.

\end{document}